\documentclass[a4paper,11pt]{scrartcl}
\usepackage[bookmarks,colorlinks,breaklinks]{hyperref}  
\hypersetup{linkcolor=blue,citecolor=blue,filecolor=dullmagenta,urlcolor=darkblue} 
\usepackage{amsmath, amssymb, amsbsy, amsthm}

\newtheorem{theorem}{Theorem}[section]
\newtheorem{lemma}[theorem]{Lemma}
\newtheorem{cor}[theorem]{Corollary}
\newtheorem{proposition}[theorem]{Proposition}
\theoremstyle{definition}

\theoremstyle{remark}

\newtheorem{remark}[theorem]{Remark}

\numberwithin{equation}{section}

\usepackage{graphicx}
\usepackage{slashed} 
\usepackage{comment}
\usepackage{bbm}

\usepackage[T1]{fontenc}
\usepackage[latin1]{inputenc}
\usepackage{dsfont}

\def\beq{\begin{equation}}
\def\eeq{\end{equation}}
\def\bea{\begin{eqnarray}}
\def\eea{\end{eqnarray}}

\newcommand{\Tr}{\,{\rm Tr}\,}
\def\={\ =\ }

\newcommand{\dd}{\ensuremath{\mathrm{d}}}

\newcommand{\e}{\ensuremath{\,\mathrm{e}\,}}
\renewcommand{\i}{\ensuremath{\,\mathrm{i}\,}}
\newcommand{\N}{\ensuremath{\mathbb{N}}}
\newcommand{\R}{\ensuremath{\mathbb{R}}}
\newcommand{\C}{\ensuremath{\mathbb{C}}}
\newcommand{\Z}{\ensuremath{\mathbb{Z}}}
\newcommand{\GS}{\ensuremath{\scal^\alpha_\alpha}}

\newcommand{\grad}{\ensuremath{\mathrm{\grad}\,}}

\newcommand{\p}[1]{\partial_{#1}}

\newcommand{\de}[2]{\ensuremath{{\frac{d #1}{d #2}}}}

\newcommand{\ket}[1]{|#1\rangle}
\newcommand{\bra}[1]{\langle #1|}
\newcommand{\bk}[2]{\langle #1|#2\rangle}
\newcommand{\kb}[2]{|#1\rangle\langle#2|}
\newcommand{\bkl}[2]{\left\langle\left. #1\right|#2\right\rangle}

\newcommand{\ep}[1]{\epsilon_{#1}}

\newcommand{\eps}{\mathcal{E}}
\newcommand{\f}{\mathcal{F}}
\newcommand{\scal}{\mathcal{S}}
\newcommand{\mcal}{\mathcal{M}}
\renewcommand{\u}{\mathbbm{1}}
\renewcommand{\d}{\ensuremath{{\sf D}}}
\renewcommand{\de}{\ensuremath{{\sf D}^2_{\rm E}}}
\newcommand{\di}[1]{\ensuremath{{\sf D}}_{#1}}
\newcommand{\dt}{\ensuremath{\tilde{\sf D}}}

\newcommand{\dti}[1]{\ensuremath{\tilde{\sf D}}_{#1}}
\renewcommand{\v}[1]{\ensuremath{\vech{V}_{#1}}}
\newcommand{\h}{\ensuremath{\vech{H}}}
\newcommand{\ho}{\ensuremath{\vech{H}_{\rm osc}}}
\renewcommand{\a}{\ensuremath{\vech{a}}}

\renewcommand{\k}{\ensuremath{{\sf K}}}
\newcommand{\kt}{\ensuremath{\tilde{\sf K}}}
\newcommand{\sigmat}{\ensuremath{(1-\sigma)}}

\def\appendix#1{\addtocounter{section}{1}\setcounter{equation}{0}
\renewcommand{\thesection}{\Alph{section}}
\section*{Appendix \thesection\protect\indent \parbox[t]{9.715cm} {#1}}
\addcontentsline{toc}{section}{Appendix \thesection\ \ \ #1} }

\newcommand{\eq}{\begin{eqnarray}}
\newcommand{\eqend}{\end{eqnarray}}

\def\vec#1{\mathchoice{\mbox{\boldmath$\displaystyle#1$}}{\mbox{\boldmath$\textstyle#1$}}{\mbox{\boldmath$\scriptstyle#1$}}{\mbox{\boldmath$\scriptscriptstyle#1$}}}
\newcommand{\vech}[1]{\ensuremath{\vec{\hat{#1}}}}
\newcommand{\vect}[1]{\ensuremath{\vec{\tilde{#1}}}}
\newcommand{\we}[1]{\ensuremath{\hat{\mathcal{W}}[#1]}}
\newcommand{\wi}[1]{\ensuremath{{\sf W}}[#1]}

\title{Duality covariant quantum field theory on noncommutative
  Minkowski space}

\author{Andr\'e Fischer\thanks{Email: {\tt 
    afischer@itp.uni-hannover.de}}\\[0.2cm]
  {\normalsize\slshape Institut f\"ur Theoretische Physik}\\[-0.2cm]
  {\normalsize\slshape Universit\"at Hannover}\\[-0.2cm]
  {\normalsize\slshape Appelstra\ss e 2, D-30167 Hannover,
    Germany}\\[+0.5cm] 
Richard J. Szabo\footnote{Email: {\tt 
    R.J.Szabo@ma.hw.ac.uk}}\\[0.2cm]
  {\normalsize\slshape Department of Mathematics and Maxwell Institute
    for Mathematical Sciences}\\[-0.2cm]
  {\normalsize\slshape Heriot-Watt University}\\[-0.2cm]
  {\normalsize\slshape Colin Maclaurin Building, Riccarton, Edinburgh
    EH14 4AS, U.K.}
}

\begin{document}

\maketitle

\begin{abstract}
We prove that a scalar quantum field theory defined on noncommutative
Minkowski spacetime with noncommuting momentum coordinates is
covariant with respect to the UV/IR duality which exchanges
coordinates and momenta. The proof is based on suitable resonance
expansions of charged noncommutative scalar fields in a background
electric field, which yields an effective description of the field
theory in terms of a coupled complex two-matrix model. The two
independent matrix degrees of freedom ensure unitarity and manifest
$\vec{C\,T}$-invariance of the field theory. The formalism describes
an analytic continuation of the renormalizable Grosse-Wulkenhaar
models to Minkowski signature.
\end{abstract}

\section{Introduction}

The renormalization of noncommutative quantum field theories has
undergone enormous progress during the last few years (see
e.g.~\cite{riv07a,riv07b} for an overview). The
mixing of ultraviolet and infrared scales prohibits the successful
application of conventional renormalization schemes, such as the
Wilsonian approach~\cite{min00}. Grosse and Wulkenhaar understood the
appearance of UV/IR mixing in scalar $\phi_{2d}^{\star 4}$ theory as an
anomaly due to a missing marginal term in the
Lagrangian~\cite{gw03,gw05}. A certain UV/IR duality symmetry of the
theory under symplectic Fourier transformation of the
fields~\cite{ls02} eliminates UV/IR mixing. In order to make their 
propagator covariant under this duality, they added a harmonic
oscillator potential to the free Lagrangian. The analysis of Grosse
and Wulkenhaar has been successfully extended to a variety of other
models~\cite{gmrvt06,vt06a,lsz03,lsz04,gs06a,gs06b,ww08,gr08}, and it
is believed that a constructive definition of these quantum field
theories may be possible due to the absence of
renormalons~\cite{dgmr07,lvtw07,ggr08}. The UV/IR 
duality has been recently interpreted in terms of metaplectic
representations of the Heisenberg group in~\cite{bgr08}, where the
analog of the Grosse-Wulkenhaar model has also been defined on
solvable symmetric spaces.

The duality covariant propagators in the original field theories
studied in~\cite{ls02} govern the propagation of charged scalar
particles in a constant magnetic background. Heuristically, the
duality exchanges infrared and ultraviolet divergences, such that both
divergences can be cut off simultaneously. This enables the standard
Wilsonian renormalization procedure to be properly applied. However,
thus far all models considered have been formulated in Euclidean
space. In this paper we will investigate how the duality covariant
scalar quantum field theories are modified in Minkowski space with
maximal rank noncommutativity.

In contrast to the commutative case, the perturbative
dynamics of noncommutative field theories in Minkowski signature
cannot be simply obtained via a Wick rotation of their Euclidean
counterparts~\cite{BDFP02,Bahns04,ry03,ls02a}. In non-planar graphs,
the Heaviside function implementing time-ordering and the two-point
function cannot be combined to yield twisted convolution products
of Feynman propagators. A careful analysis treating both functions on
a different footing reveals that the renormalization properties in
Minkowski signature are very different than on Euclidean
space~\cite{Bahns04}, and it has been suggested that the UV/IR
mixing problem may be far less severe or even absent in this case.

In order to analyse the UV/IR duality in Minkowski signature, we will
continue the models investigated in~\cite{ls02,lsz03,lsz04} to
Minkowski space. Thus we will consider a complex scalar field in a
background \emph{electric} field. We will establish the duality
covariance of the interacting noncommutative quantum field theory. In
doing so we will introduce a matrix basis for the expansion of fields,
which can be considered as the Minkowskian analog of the expansion in
Landau wavefunctions on noncommutative Euclidean space. The matrix
basis is the key setting for application of the Wilson-Polchinski
renormalization group equation in the Grosse-Wulkenhaar model. In
contrast to the Euclidean case, however, the Lorentzian duality
covariant field theory requires \emph{two} coupled complex matrices in
its representation as a matrix model, a necessary unitary and causal
property which does not follow by a simple Wick rotation. The
two-matrix model naturally ensures the stability and
$\vec{C\,T}$-invariance of the field 
theory. This model can thus be regarded as an analytical continuation
of the Grosse-Wulkenhaar models to noncommutative Minkowski space, and
is the starting point for the renormalization of noncommutative
quantum field theory in Lorentzian signature.

The $1+1$-dimensional Klein-Gordon operator appearing in the free part
of the duality covariant action is a special representation of the
quantum \emph{inverted} harmonic oscillator defined by the Hamiltonian
\begin{eqnarray}
 \vech{H}=\mbox{$\frac{1}{2}$}\,\big(\vech{P}^2-\omega^2\,
\vech{Q}^2\big)
\label{invho1}\end{eqnarray}
with $\omega\in\R$, 
where the position and momentum operators $\vech{Q}$ and $\vech{P}$
obey the canonical commutation relation $[\vech{Q},\vech{P}]=\i$. The
inverted harmonic oscillator emerges if one inserts an imaginary
frequency $\pm\i\omega$ into the usual quantum harmonic oscillator. As
we will see below, we can also obtain one system from the other by a
{complex scaling}. However, the spectral properties of these two
systems are completely different. Unlike the quantum harmonic
oscillator, which has a discrete spectrum bounded from below, the
inverted oscillator exhibits a continuous spectrum which is not
bounded from below. Intriguingly, even though the operator $\vech{H}$
is selfadjoint, it possesses a second set of generalized
eigenfunctions corresponding to imaginary eigenvalues. These functions
occur as residues of the original eigenfunctions analytically
continued to the complex energy plane. Such functions are well known in
the literature and are used to describe resonant states, often called
{Gamow states} (see e.g.~\cite{gad04} for a review). To uncover
these states we have to close the contour of integration over the
eigenfunction expansion in the upper or lower complex half-plane, and
the resulting discrete expansion is analogous to the expansion in
Landau wavefunctions.

From a technical standpoint, the matrix basis is derived from an
application of the Gel'fand-Maurin spectral theorem and an appropriate
resonance expansion of fields. This expansion requires truncation of
the configuration space of the field theory to a dense subspace, which
we describe in detail. Thus the integration domain for the functional
integral must be truncated, which may be thought of as an ingredient
of the duality covariant regularization of the quantum field
theory. We work in the framework of generalized functions and
Gel'fand-Shilov spaces~\cite{gel64}, which are subalgebras of Schwartz
space closed under Fourier transformation and allow for the
appropriate expansions in terms of harmonic oscillator
wavefunctions~\cite{L-CP07}. The Gel'fand-Shilov spaces are also
closed under multiplication with the noncommutative star
product~\cite{Sol07a,cmtv07,Sol07b}, and are thus natural candidates
for the configuration spaces of duality covariant noncommutative field
theories. These functional analytic techniques should all prove useful
for further development of the renormalization programme on
noncommutative Minkowski space.

The outline of the remainder of this paper is as follows. In
Section~\ref{Formulation} we give a precise formulation of the
noncommutative quantum field theory in $1+1$-dimensions and state its
duality symmetries. In Section~\ref{Quantduality} we develop in detail
the resonance expansion of our noncommutative fields and use it to
prove the duality covariance of the Lorentzian quantum field
theory. In Section~\ref{time} we describe both physical and analytic
properties of the subspace of Schwartz space on which our resonance
expansions are valid. In Section~\ref{2matrixmodel} we describe the
equivalent two-matrix model which governs the dynamics of the
duality covariant quantum field theory. In Section~\ref{HigherD} we
describe the generalization of our results to higher-dimensional
noncommutative Minkowski space. In Section~\ref{Conclusions} we
summarize our findings and discuss the prospects of using our analysis
in further directions. Finally, two appendices at the end of the paper
contain some of the more technical aspects of our development. In
Appendix~A we describe properties and the explicit analytic forms of
the generalized eigenfunctions which are used to derive the resonance
expansions. In Appendix~B we derive the explicit expression for the
free two-point Green's function in the duality covariant quantum field
theory. 

\section{Formulation of the duality covariant
  field theory\label{Formulation}}

In this section we will describe the scalar field theory we shall work
with and its duality symmetries. Let us begin by giving a heuristic
motivation behind the duality. Consider the noncommutative field
theory of a complex scalar field $\phi(\vec{x})$ in
$D$-dimensional spacetime. The noncommutativity parameters are
specified by a real constant $D\times D$ antisymmetric matrix
$\vec{\theta}$. The infrared dynamics of the quantum
field theory are mediated through the interactions of noncommutative
``dipoles''~\cite{Rey02}, which are extended degrees of freedom (rigid
``rods'') whose lengths are proportional to their transverse
momentum. For a dipole of momentum $\vec{k}$, its dipole moment is
$\vec{\theta}\cdot\vec{k}$ and the position coordinate $\vec{x}$ of
the scalar field is Bopp shifted to the commutative variable
\beq
\vec{r}=\vec{x}+\vec{\theta}\cdot\vec{k} \ .
\label{IRcoord}\eeq
The dipole degrees of freedom are created by the
operators~\cite{Rey02,VanR01}
\beq
W_{\vec{k}}[\phi]\=\Tr\exp\big(\i|\vec{k}|\,\phi(\vec{x})\big)\=
\Tr\exp\big(\i|\vec{k}|\,\phi(\vec{r}-\i\vec{\theta}\cdot
\nabla_{\vec{r}})\big) \ .
\label{openWilson}\eeq
In the case of noncommutative gauge theory, an alternative
interpretation of the infrared dynamics as a non-renormalizable
gravitational sector has been given recently in~\cite{Stein07,gsw08}.

On the other hand, the ultraviolet dynamics are governed by the
elementary quantum fields $\phi$, which create pointlike quanta of
momenta $\vec{k}$. The ultraviolet and infrared degrees of freedom are
``dual'' to one another~\cite{Rey02}. The UV/IR mixing problem can in
this way be understood as a mismatch between the dressed coordinates
(\ref{IRcoord}) and the elementary momenta $\vec{k}$. We will cure
this problem by making the UV/IR ``duality'' symmetric via
substitution of the generalized momenta
\beq
\vec{k}~\longmapsto~\vec{k}+\vec{E}\cdot\vec{x} \ ,
\label{momsubs}\eeq
where the real constant $D\times D$ antisymmetric matrix $\vec{E}$ can be
interpreted as an ``electromagnetic'' background.

For this, consider the quantum field theory of a massive, complex
scalar field $\phi(\vec{x})$ minimally coupled to a constant
electromagnetic field in flat Minkowski spacetime, and in the
background of an inverted harmonic oscillator potential. To simplify
the presentation we will focus mainly on the case of $D=1+1$
dimensions, commenting later on the extension to generic spacetime
dimension (see Section~\ref{HigherD}). The spacetime coordinates are
denoted by $\vec{x}=(t,x)=(x^\mu)$. We will denote by
$(\text{G}_{\mu\nu})=\text{diag}(1,1)$ the flat Euclidean metric and with
$(\eta_{\mu\nu})=\text{diag}(1,-1)$ the flat Minkowski metric. The
electric field strength tensor is denoted $\vec{E}=(E_{\mu\nu})$ and
\begin{eqnarray}
\f[\phi](\vec{k})\=\frac{1}{2\pi}\,\int_{\R^2}\,\dd\vec{x}~
\e^{-\i \vec{k}\cdot\vec{x}}\,\phi(\vec{x})
\qquad \mbox{with} \quad 
\vec{k}\cdot\vec{x}\=k_\mu\,x^\mu~:=~\eta_{\mu\nu}\,k^\mu \,x^\nu
\label{Fourier}\end{eqnarray}
is the usual Fourier transformation of the field $\phi(\vec{x})$.

The field theory is defined by the action $S=S_0+g^2\,S_{\rm int}$
with the free part given by 
\begin{eqnarray}
S_0&=&\int_{\R^2}\, \dd\vec{x}~\phi^*(\vec{x})\,
\big(\sigma\,\d^2+\sigmat\,\dt^2+\mu^2\big)\phi(\vec{x}) \ ,
\label{freepart}\end{eqnarray}
where the parameter $\sigma\in[0,1]$, $\mu^2>0$ is the mass parameter,
and $\d^2=\eta^{\mu\nu}\,\di{\mu}\,\di{\nu}$ with $\d_\mu$ the
generalized momentum operators defined as
\begin{eqnarray}
\di{\mu}=\mbox{$\frac{1}{\sqrt{2}}$}\,\left(-\i\p{\mu}+E_{\mu\nu}\,
x^\nu\right) \ ,
\label{genmomops}\end{eqnarray}
and $(\p{\mu})=(\partial/\partial x^\mu)=(\p{t},\p{x})$. The 
generalized momenta obey the commutation relations
\begin{eqnarray}
[\di{\mu},\di{\nu}]=\i E_{\mu\nu} \ ,
\end{eqnarray}
which allows us to interpret the constant electric field strength 
$E_{\mu\nu}=E\,\epsilon_{\mu\nu}$ as a parameter which produces
noncommuting momentum space coordinates. The other kinetic operator
$\dt^2=\eta^{\mu\nu}\,\dti{\mu}\,\dti{\nu}$ is specified in terms of
the ``dual'' momenta
\begin{eqnarray}
 \dti{\mu}=\mbox{$\frac{1}{\sqrt{2}}$}\,\left(-\i\p{\mu}-E_{\mu\nu}\,
x^\nu\right)
\end{eqnarray}
which commute with the operators $\di{\mu}$ and are obtained from
(\ref{genmomops}) by the charge conjugation transformation
$\vec{C}:E_{\mu\nu}\mapsto-E_{\mu\nu}$.

The interaction part consists of the two inequivalent, noncommutative
quartic interactions
\begin{eqnarray}
 S_{\rm int}&=&\int_{\R^2}\, \dd\vec{x}~\big[\alpha\,(
\phi^*\star\phi\star\phi^*\star\phi)(\vec{x})+\beta\,(
\phi^*\star\phi^*\star\phi\star\phi)(\vec{x})\big]
\label{intpart}\end{eqnarray}
weighted by the real parameters $\alpha$ and $\beta$. We will use the
usual Gr\"onewold-Moyal star-product which may be defined by the
twisted convolution product
\begin{eqnarray}
f(\vec{x})\star g(\vec{x})&=&\frac1{(2\pi)^2}\,\int_{\R^2}\,\dd\vec{k}~
\int_{\R^2}\,\dd\vec{p}~\f[f](\vec{k})\,\f[g](\vec{p})~
\e^{\frac{\i}{2}\,\theta\,\ep{\mu\nu}\,k^{\mu}\,p^{\nu}}~\e^{\i(\vec{k}+
\vec{p})\cdot\vec{x}} \nonumber\\[4pt] &=& 
\frac{1}{(2\pi\,\theta)^2}\,\int_{\R^2}\,\dd\vec{x}_1~\int_{\R^2}\,
\dd\vec{x}_2~f(\vec{x}_1)\,g(\vec{x}_2)~\e^{-\frac{2\i}\theta\,
\ep{\mu\nu}\,(\vec{x}_1-\vec{x})^\mu\,(\vec{x}_2-\vec{x})^\nu} \ .
\label{starprod}\end{eqnarray}
We assume here that $\phi\in\scal(\R^2)$ 
is a Schwartz test function on $\R^2$ for
simplicity. The Fourier transformation (\ref{Fourier}) is a
topological automorphism of Schwartz space, and the twist factor
$\exp(\frac{\i}{2}\,\theta\,\ep{\mu\nu}\,k^{\mu}\,p^{\nu})$ is a
multiplier for this space. Later on we will further restrict this
space to an appropriate subspace.

We are now ready to give a precise formulation of the duality in the
classical field theory.
\begin{theorem}\label{thm1}
The action
\begin{eqnarray}
S\=S_0+g^2\,S_{\rm int}~=:~S[\phi;\vec{E},g,\vec{\theta}]
\end{eqnarray}
defined above obeys
\begin{eqnarray}
S[\phi;\vec{E},g,\vec{\theta}]=S\big[\tilde{\phi}\,;\,\vec{E}\,,\,
\tilde{g}\,,\,\vect{\theta}\,\big]\,,
\end{eqnarray}
where
\begin{eqnarray}
\tilde{\phi}(\vec{x})=\sqrt{\big|\det(\vec{E})
\big|}~\f[\phi](\vec{E}\cdot \vec{x})
\end{eqnarray}
and $\f[\phi](\vec{k})$ is the Fourier transform of
$\phi(\vec{x})$. The transformed coupling parameters are
\begin{eqnarray}
\vect{\theta}\=-4\vec{E}^{-1}\,\vec{\theta}^{-1}\,\vec{E}^{-1}
\qquad \mbox{and} \qquad
\tilde{g}\=2\big|\det(\vec{E}\cdot\vec{\theta})\big|^{-1/2}\,g \ .
\end{eqnarray}
Moreover, the transformation $(\phi;\vec{E},g,\vec{\theta})\mapsto
(\tilde{\phi};\vec{E},\tilde{g},\vect{\theta}\,)$ is a duality of the
field theory, i.e. it generates a cyclic group of order two. 
\end{theorem}
At the special points $\theta=\pm\,2/E$ the field theory is completely
invariant under Fourier transformation (up to the sign of $\theta$),
and it is said to be
\emph{self-dual}. The proof of Theorem~\ref{thm1} is identical to that
of~\cite[Prop.~1]{ls02}, which holds irrespectively of the signature of
the spacetime metric. As in~\cite{ls02}, each of the differential
operators $\d^2$ and $\dt^2$ is invariant under Fourier transformation
up to a rescaling. The duality covers both cases $\sigma=0$ and 
$\sigma=1$ representing charged scalar fields in a background electric
field alone, analogously to the Euclidean models of~\cite{ls02}. Since
\beq
\d^2+\dt^2=-\partial^\mu\,\partial_\mu-E^2\,x^\mu\, x_\mu \ ,
\label{D2Dt2}\eeq
the choice $\sigma=\frac12$ corresponds to scalar fields
in an inverted harmonic oscillator potential alone and is closest to
the conventional field theories on noncommutative Minkowski space with
no background electric field. In the Euclidean setting it is this
choice which renders the standard noncommutative $\phi^4$-theory
renormalizable to all orders of perturbation theory by giving the free
propagator the necessary decay behaviour for a multiscale
slicing~\cite{gw03}, achieved by discretization of the spectrum of the
free Hamiltonian via the effective infrared regularization provided by
the confining harmonic oscillator potential.

Let us now turn to the duality at the full quantum level in Minkowski
spacetime. Formally, the quantum field theory defined by the classical
action above is duality invariant even for Minkowski metric. It is
defined by the usual perturbative, formal functional integral
\begin{eqnarray}
 Z[J]=\int \,\mathcal{D}\phi~\mathcal{D}\phi^*~
\exp\big(
\i S[\phi;\vec{E},g,\vec{\theta}]+\i\langle\phi,J\rangle+\i\langle
  J,\phi\rangle\big)
\label{ZJdef}\end{eqnarray}
where $\langle f,g\rangle:=\int_{\R^2}\,
\dd\vec{x}~f^*(\vec{x})\,g(\vec{x})$, with independent external
sources $J(\vec{x})$ and $J^*(\vec{x})$. The generating
functional of all connected Green's functions is given by
\begin{eqnarray}
 \mathcal{G}[J]\=-\log\frac{Z[J]}{Z[0]}~=:~
\mathcal{G}[J;\vec{E},g,\vec{\theta}]\,.
\end{eqnarray}
As in~\cite{ls02}, due to the duality covariance of the classical
action $S$, the invariance of the functional integration measure under
the transformation $\phi\mapsto\tilde\phi$, and the fact that $\langle
\phi,J\rangle=\langle\tilde{\phi},\tilde{J}\,\rangle$, we formally
obtain the identity
\begin{eqnarray}
 \mathcal{G}[J;\vec{E},g,\vec{\theta}]=\mathcal{G}\big[\tilde
 J\,;\,\vec{E}\,,\,g\,,\,\vect{\theta}\,\big] \ .
\end{eqnarray}
However, a proper treatment requires a specification of ultraviolet
and infrared regularizations. As we will show in the next section,
there exists a duality invariant regularization which cures all
possible divergences of the quantum field theory. Assuming this
is properly done we then see that the regularized quantum field theory
is duality invariant.

We summarize this result as follows.
\begin{theorem}\label{thm2}
There exists a regularization which is
invariant under the duality transformation given in
Theorem~\ref{thm1}. Moreover, with this regularization and for
Minkowski spacetime metric, all Feynman amplitudes of the quantum field
theory are convergent. The corresponding regularized generating
functional $\mathcal{G}_\Lambda$ of all connected Green's functions,
where $\Lambda$ is a cut-off parameter defined by the regularization,
is therefore well-defined. It possesses the duality symmetry
\begin{eqnarray}
\mathcal{G}_\Lambda[J;\vec{E},g,\vec{\theta}]=\mathcal{G}_\Lambda\big[
\tilde{J}\,;\,\vec{E}\,,\,\tilde{g}\,,\,\vec{\tilde{\theta}}\,\big]
\end{eqnarray}
where
$\tilde{J}(\vec{x})=\sqrt{|\det(\vec{E})|}~\f[J](\vec{E}\cdot\vec{x})$
and $\f[J]$ is the Fourier transform of $J$.
\end{theorem}

The key ingredient for the quantum duality is the existence of a
regularization of the quantum field theory which respects the
duality. In~\cite{ls02} it was shown that in Euclidean space there
exists a natural regularization for the theory. Rather than expanding
the fields in plane waves, it is more natural to expand them in
eigenfunctions of the Landau Hamiltonian
$\de:=\text{G}^{\mu\nu}\,\di{\mu}\,\di{\nu}$ where $\text{G}^{\mu\nu}$
is the Euclidean metric, which diagonalizes the free part of the
action. Since the Landau wavefunctions are not eigenfunctions of the
operator $\d^2=\eta^{\mu\nu}\,\di{\mu}\,\di{\nu}$, the proof given
in~\cite{ls02} does not directly apply in Minkowski spacetime. In the
next section we will develop an analogous expansion for Minkowski
signature, which will allow us to prove the theorem in a similar
manner. The technical details are rather intricate in this case, and
we will uncover some surprising differences from the Euclidean case.

In what follows it will be useful to employ the Weyl-Wigner
correspondence of noncommutative field theory~\cite{sza01}. It will
play a central role in our analysis of both the free action where no
noncommutativity shows up and in our analysis of the noncommutative
interactions. In the former case it will allow us to switch easily
between different representations of our Klein-Gordon operator $\d^2$,
while on the other hand we can utilize some nice properties of this
mapping to give explicit expressions for the generalized
eigenfunctions of $\d^2$ in our original representation and thus prove
the duality covariance of the model. In the latter case we can use the
same property to map our quantum field theory onto a matrix model.

The Weyl-Wigner correspondence provides a one-to-one correspondence 
between the algebra of fields on $\R^2$ and a ring of operators with
(suitably normalized) trace $\Tr$, constructed through replacing the
local coordinates $x^\mu$ of $\R^2$ by Hermitean operators
$\vech{x}^\mu$ obeying the Heisenberg commutation relations 
\begin{eqnarray}
\big[\vech{x}^\mu\,,\,\vech{x}^\nu\big]=\i\theta^{\mu\nu} \ .
\label{comm1}\end{eqnarray}
Given a Schwartz function $f(\vec{x})$, we introduce its {Weyl
  symbol}
\begin{eqnarray}
 \we{f}=\frac1{2\pi}\,\int_{\R^2}\,\dd\vec{k}~\f[f](\vec{k})\,
\exp\big(\i k_\mu\,\vech{x}^\mu\big) \ ,
\end{eqnarray}
which is a compact operator. The transformation
$f(\vec{x})\mapsto\we{f}$ is invertible with inverse given
by~\cite{sza01}
\begin{eqnarray}
 f(\vec{x})\=\frac1{2\pi}\,\int_{\R^2}\,\dd\vec{k}~\e^{-\i
  \vec{k}\cdot\vec{x}}\,\Tr\big(\we{f}\,\exp(\i
k_\mu\,\vech{x}^\mu)\big) ~=:~\wi{\we{f}}(\vec{x})
\ ,
\end{eqnarray}
which is often called the Wigner distribution function of the operator
$\we{f}$. One has~\cite{sza01}
\begin{eqnarray}
\we{f}\,\we{g}\=\we{f\star g} \qquad \mbox{and} \qquad
\wi{\vech{f}\,}\star\wi{\vech{g}}\=\wi{\vech{f}\,\vech{g}}
\label{wewistar}\end{eqnarray}
for arbitrary Schwartz functions $f(\vec{x})$, $g(\vec{x})$ and
compact operators $\vech{f}$, $\vech{g}$, while
\beq
\int_{\R^2}\,\dd\vec{x}~f(\vec{x})\=\Tr\big(\we{f}\big)
\qquad \mbox{and} \qquad
\Tr\big(\vech{f}\,\big)\=\int_{\R^2}\,\dd\vec{x}~
\wi{\vech{f}\,}(\vec{x}) \ .
\label{wewiintTr}\eeq

\section{Quantum duality on noncommutative Minkowski
  space\label{Quantduality}} 

In order to prove Theorem \ref{thm2} for Minkowski spacetime, we will
show how to expand the boson fields in a discrete set of generalized
eigenfunctions of the operator $\d^2$. One of the most important facts
needed for the proof in the Euclidean case is that the operator $\de$
has a discrete spectrum. By expanding the fields in this discrete
basis of eigenfunctions, the field theory can be mapped onto a
matrix model and regularized by cutting off the sums appearing in the 
Feynman amplitudes at some finite matrix rank $N$. The Minkowski case
is much more subtle, since the spectrum of the operator $\d^2$ is the
whole real line $\R$. Nevertheless, we will show that there exists an
appropriate space of fields in which a discrete expansion is
possible. This will enable us to apply the arguments given in the
Euclidean case. In the following we will extend and generalize
some results of~\cite{chr02} which were
obtained in a different context than ours.

\subsection{Mapping onto the inverted harmonic
  oscillator\label{InvHO}} 

To analyse the duality invariance of our model it is necessary to fix
the self-dual point $\theta^{\mu\nu}=\theta\,\epsilon^{\mu\nu}$ with
$\theta=2/E$. (When we study the interacting field theory later on, we
will assume that $\theta$ and $E$ are independent parameters.)
\begin{lemma}
There exists a classical Hamiltonian
$H(\vec{x})=\mbox{$\frac{1}{2}$}\,\big(x^2-t^2\big)$ such that the
actions of $\d^2$ and $\dt^2$ on any function $f(\vec{x})$ is
proportional to the star product of $H$ with that function as
\begin{eqnarray}
 \d^2f(\vec{x})=E^2\,H(\vec{x})\star f(\vec{x}) \qquad \mbox{and}
 \qquad \dt^2 f(\vec{x})=E^2\,f(\vec{x})\star H(\vec{x}) \ .
\label{dh}\end{eqnarray}
\label{HOlemma}\end{lemma}
\begin{proof}
The first equality follows from an elementary calculation
\begin{eqnarray}
 \mbox{$\frac{1}{2}$}\,\big(x^2-t^2\big)\star
 f(\vec{x})&=&\mbox{$\frac{1}{2}$}\,
\left[x^2-t^2 -
  2\i(\theta/2)\,(x\,\p{t}+t\,\p{x})-(\theta/2)^2\,(\p{t}^2-\p{x}^2)
\right]f(\vec{x}) \\[4pt]
&=&\mbox{$\frac{1}{2E^2}$}\,\left(-\partial^\mu\,\partial_\mu-2\i E\,
\ep{\mu\nu}\,x^\nu\,\partial^\mu-E^2\,x^\mu\, x_\mu\right)f(\vec{x})\=
\mbox{$\frac{1}{E^2}$}\,\d^2 f(\vec{x}) \ , \nonumber 
\end{eqnarray}
where in the second line we have set $\theta=2/E$. An analogous
calculation establishes the second equality in (\ref{dh}).
\end{proof}

Instead of the operators $\d^2$ and $\dt^2$, we may thus work with the
classical Hamiltonian $H(\vec{x})$, or even better with its Weyl
symbol $\we{H}=:\vech{H}$. This operator is given by
\begin{eqnarray}
  \vech{H}\=\mbox{$\frac{1}{2}$}\,\big(\we{x}^2-\we{t}^2\big)~=~
\mbox{$\frac{1}{2}$}\,\big(\vech{p}^2-\vech{q}^2\big) \ ,
\end{eqnarray}
where the operators $\vech{p}:=\we{x}$ and $\vech{q}:=\we{t}$ obey the
commutation relation
\begin{eqnarray}
\big[\vech{q}\,,\,\vech{p}\big]\=\we{t\star x-x\star
  t}\=\i\theta\=2\i/E \ .
\end{eqnarray}
The operator $\vech{H}$ is known as the inverted harmonic
oscillator Hamiltonian. Its spectral properties are reviewed below.

\subsection{Rigged Hilbert space and resonance
  expansion}\label{resexp1}

Resonance states were first introduced to describe decay phenomena in
nuclei. They correspond to complex energy eigenvalues of a
Hamiltonian. The mathematical object in which to embed such states is
a {rigged Hilbert space}. The extension of the usual Hilbert
space to a rigged Hilbert space is also necessary to deal with
continuous spectra of selfadjoint operators. The spectral theorem for
Hilbert spaces, which is only valid for operators with discrete
spectra, can be extended to these operators by the {Gel'fand-Maurin
  theorem} (also known as the {nuclear spectral theorem}). See
e.g.~\cite{dlm05} and references therein for an introduction to rigged 
Hilbert spaces in quantum mechanics.

A rigged Hilbert space is roughly speaking a triplet of spaces 
\begin{eqnarray}
\Phi~\subset~\mathcal{H}~\subset~\Phi' \ ,
\end{eqnarray}
where $\Phi$ is a dense, topological vector subspace of an
infinite-dimensional Hilbert space $\mathcal{H}$ and $\Phi'$ is
its topological dual, i.e. the space of continuous linear functionals
on $\Phi$. The action of a functional ${F}\in\Phi'$ on a vector
${\phi}\in\Phi$ will be denoted $\bk{\phi}{F}\in\C$. It is the
extension of the inner product on $\mathcal{H}$ to
$\Phi\times\Phi'$. If $\vech{A}$ is a selfadjoint operator on
$\mathcal{H}$, then a complex number $\lambda\in\C$ is called a
\emph{generalized eigenvalue} of $\vech{A}$ if there is a nonzero
functional ${F_\lambda}\in\Phi'$, called a \emph{generalized
  eigenvector}, such that for any ${\phi}\in\Phi$ one has
\begin{eqnarray}
\bk{\phi}{\vech{A}F_\lambda}~:=~\bk{\vech{A}\phi}{F_\lambda}
\=\lambda\,\bk{\phi}{F_\lambda} \ .
\end{eqnarray}
In this way the operator $\vech{A}$ can be extended to the dual
space $\Phi'$, and it is possible to make sense of complex
eigenvalues of selfadjoint operators.

By the Gel'fand-Maurin theorem, for every selfadjoint operator
$\vech{A}$ there exists a measure $\dd\mu$ on the spectrum
$\Sigma(\vech{A})\subset\R$, which for an absolutely continuous
spectrum can be chosen to be Lebesgue measure, such that for almost
every $\lambda\in\Sigma(\vech{A})$ we can find a nonzero functional
$\ket{F_\lambda}\in\Phi'$ with 
\begin{eqnarray}
\vech{A}\ket{F_\lambda}=\lambda\,\ket{F_\lambda} \ .
\end{eqnarray}
These generalized eigenvectors cover the spectrum and form a complete
set, and thus for an arbitrary vector $\ket{\phi}\in\Phi$ provide the
decomposition
\begin{eqnarray}
\ket{\phi}=\sum_{\lambda_n\in\Sigma_p(\vech{A})}\,
\bk{F_{\lambda_n}}{\phi}\,\ket{F_{\lambda_n}}+
\int_{\Sigma_c(\vech{A})}\,\dd\lambda~\bk{F_\lambda}{\phi}\,\ket{F_\lambda}
\end{eqnarray}
where $\Sigma_p(\vech{A})$ and $\Sigma_c(\vech{A})$, with
$\Sigma(\vech{A})=\Sigma_p(\vech{A})\cup\Sigma_c(\vech{A})$, are
respectively the point and continuous spectrum of $\vech{A}$. On the
domain $\Phi$, the Gel'fand-Maurin theorem allows the spectral
representation for $\vech{A}$ given by
\begin{eqnarray}
\vech{A}\,\big|_\Phi=\sum_{\lambda_n\in\Sigma_p(\vech{A})}\,\lambda_n\,
\ket{F_{\lambda_n}}\bra{F_{\lambda_n}}+\int_{\Sigma_c(\vech{A})}\,
\dd\lambda~\lambda\,\ket{F_{\lambda}}\bra{F_{\lambda}} \ .
\end{eqnarray}
We will now investigate the spectral structure and the rigged Hilbert
space of the inverted harmonic oscillator Hamiltonian $\h$. The
spectral properties of $\h$ were analysed in~\cite{chr03,chr04}.

Our first goal is to find the eigenfunctions of
$\vech{H}$ and determine the rigged Hilbert space in which an
eigenvector expansion is possible. The spectrum of $\h$ is $\R$ and
the rigged Hilbert space is given by 
\begin{eqnarray}
\mathcal{S}(\R)~\subset~ L^2(\R)~\subset~ \mathcal{S}'(\R) \ ,
\label{IHOrigged}\end{eqnarray}
where $\mathcal{S}(\R)$ is the Schwartz space and $\mathcal{S}'(\R)$
is the dual space of tempered distributions. We will then show that there
exists a set of generalized eigenfunctions corresponding to imaginary
eigenvalues. Since these eigenvalues do not belong to the spectrum, we
cannot simply apply the Gel'fand-Maurin theorem to achieve a discrete
expansion on the rigged Hilbert space (\ref{IHOrigged}). Nevertheless,
it is the expansion in these eigenfunctions we are interested in. We
will show that they arise as residues of the original
eigenfunctions corresponding to the continuous eigenvalues. Through a
further restriction of the domain of the Hamiltonian $\vech{H}$, we
can apply the residue theorem to reduce the continuous eigenvector
expansion to a discrete one.

\begin{lemma}
The operator $\vech{H}$ is selfadoint on $L^2(\R)$ with spectrum
$\Sigma(\vech{H})=\R$.
\label{selfadlemma}\end{lemma}
The proof of Lemma~\ref{selfadlemma} can be found in~\cite{chr03}. As
mentioned above, since we are dealing with a continuous spectrum we
cannot expect the eigenfunctions to live in $L^2(\R)$. We will now
choose a special representation to see what the eigenfunctions of
$\vech{H}$ look like. In order to work in a similar convention
to~\cite{chr04}, we multiply $\vech{H}$ by $E'\,^2:=(E/2)^2$. Denoting
by $\ket{q}$ the eigenbasis of $\vech{q}$ with eigenvalue $q\in\R$, we
get the eigenvalue equation
\begin{eqnarray}
 \mbox{$\frac{1}{2}$}\,\big(-\p{q}^2-E'\,^2\,q^2\big)\chi_\pm^\eps(q)=
\eps\,\chi_\pm^\eps(q) \ .
\end{eqnarray}
Since the differential operator in this equation is parity invariant,
each eigenvalue $\eps$ is two-fold degenerate as indicated through the
additional index $\pm$ carried by the eigenfunctions. Substituting
$z=\sqrt{2\i E'}~q$ the eigenvalue equation can be rearranged to the
form
\begin{eqnarray}
 \left(\p{z}^2+\nu+\mbox{$\frac{1}{2}-\frac{z^2}{4}$}\right)
\chi_\pm^\eps(z)=0 \ ,
\label{paraboleq}\end{eqnarray}
where
\begin{eqnarray}
 \nu=-\i\frac{\eps}{E'}-\frac{1}{2} \ .
\label{nudef}\end{eqnarray}

The differential equation \eqref{paraboleq} is solved by 
the parabolic cylinder functions $D_\nu(z)$ which are defined by 
\begin{eqnarray}
 D_\nu(z)&=&\frac{1}{\Gamma(-\nu)}~\e^{-\frac{1}{4}\,z^2}\,
\int_0^\infty\, \dd t~\e^{-z\,t}~\e^{-\frac{1}{2}\,t^2}\,t^{-\nu-1} \ .
\label{integral_representation}\end{eqnarray}
In particular, every solution is a linear combination of the functions
$D_\nu(z)$, $D_\nu(-z)$, $D_{-\nu-1}(\i z)$ and $D_{-\nu-1}(-\i
z)$. Only two of them are linearly independent. As claimed above, the
spectrum is the entire real line and is thus not bounded from
below. This property is exactly what we need to construct our discrete
expansion.

For our purposes we will need two different sets of normalized
eigenfunctions $\chi_\pm^\eps$ and $\eta_\pm^\eps$, both corresponding
to the eigenvalue $\eps$. They are related to each other by
$\eta_\pm^\eps(q)=\chi_\pm^\eps(q)^*$, and are given
explicitly by~\cite{chr04}
\begin{eqnarray}
\chi^\eps_\pm(q)&=&\frac{C}{\sqrt{2\pi\, E'}}\,
{\i}^{\frac\nu2+\frac{1}{4}}\,\Gamma(\nu+1)\,D_{-\nu-1}\big(\mp\,
\sqrt{-2\i E'}~q\big) \ , \nonumber \\[4pt]
\eta^\eps_\pm(q)&=&\frac{C}{\sqrt{2\pi \,E'}}\,
{\i}^{\frac\nu2+\frac{1}{4}}\,\Gamma(-\nu)\,D_{\nu}\big(\mp\,\sqrt{2\i
  E'}~q\big)\label{eta}
\end{eqnarray}
where $C=(E'/2\pi^2)^{1/4}$. These functions satisfy the
orthonormality and completeness relations
\begin{eqnarray}
 \int_{\R}\,\dd
 q~\chi_\pm^{\eps_1}(q)^*\,\chi_\pm^{\eps_2}(q)
\=\delta(\eps_1-\eps_2) \qquad \mbox{and} \qquad 
\int_{\R}\,\dd\eps~\chi_\pm^{\eps}(q)^*\,\chi_\pm^{\eps}(q'\,)
\=\delta(q-q'\,) \ ,
\end{eqnarray}
and analogous relations for $\eta_\pm^{\eps}$. These \emph{generalized
  eigenfunctions} belong to the space of tempered distributions
$\mathcal{S}'(\R)$. Applying the Gel'fand-Maurin theorem to our
inverted harmonic oscillator we get two expansions for every Schwartz
function $\phi\in\mathcal{S}(\R)$ given by
\begin{eqnarray}
\phi(q)\=\sum_{s=\pm}~\int_{\R}\,
\dd\eps~\bkl{\chi^\eps_s}{\phi}\,\chi^\eps_s(q) \qquad \mbox{and}
\qquad
\phi(q)\=\sum_{s=\pm}~\int_{\R}\,\dd\eps~\bkl{\eta^\eps_s}{\phi}\,
\eta^\eps_s(q)\label{exp2} \ ,
\end{eqnarray}
and two spectral decompositions for $\h$ given by
\begin{eqnarray}
\h\=\sum_{s=\pm}~\int_{\R}\,\dd\eps~\eps\,\ket{\chi^\eps_s}
\bra{\chi^\eps_s} \qquad \mbox{and} \qquad
\h\=\sum_{s=\pm}~\int_{\R}\,\dd\eps~\eps\,\ket{\eta^\eps_s}
\bra{\eta^\eps_s} \ .
\end{eqnarray}

As mentioned before, in addition to the eigenfunctions given above the 
Hamiltonian $\h$ possesses a set of generalized eigenfunctions
corresponding to a discrete set of imaginary generalized eigenvalues
which do not appear in its spectrum. As shown in~\cite{chr04}, there
is a connection with the spectrum of the ordinary harmonic
oscillator. By introducing the Hermitean scaling operator  
\begin{eqnarray}
 \v{\lambda}:=\exp\big(\mbox{$\frac{\lambda}{2}$}\,(\vech{p}\,\vech{q}+
\vech{q}\,\vech{p})\big)
\end{eqnarray}
for $\lambda\in\R$, we can use Hadamard's lemma to compute
\begin{eqnarray}
 \v{\lambda}\,\big(\vech{p}^2-\vech{q}^2\big)\,\v{\lambda}^{-1}&=&
\e^{2\i\lambda\,\theta}\,\big(\vech{p}^2-\e^{-4\i\lambda\,\theta}\,
\vech{q}^2\big) \ .
\end{eqnarray}
Setting $\lambda=\pm\,\frac{\pi}{4\theta}$, we see that the inverted
harmonic oscillator Hamiltonian $\h$ is related to the ordinary
harmonic oscillator Hamiltonian $\ho=\frac12\,(\vech{p}^2+\vech{q}^2)$
through
\begin{eqnarray}
 \pm\i\v{\pm}\,\ho\,\v{\pm}^{-1}=\h
\end{eqnarray}
with $\v{\pm}:=\vech{V}_{\mp\,\pi/4\theta}$. This enables us to
construct two different sets of generalized eigenfunctions of
$\vech{H}$ by acting on the eigenfunctions of the harmonic oscillator
$\ket{m}$ with the operators $\v{\pm}$. This leads to
\begin{eqnarray}
 \h\ket{f_m^\pm}~:=~\h\,\v{\pm}\ket{m}\=\pm\i\v{\pm}\,\ho\ket{m}\=\pm\i 
\theta\,\big(m+\mbox{$\frac12$}\big)\,\ket{f_m^\pm} \ ,
\end{eqnarray}
where $\theta\,(m+\frac12)=(2/E)\,(m+\frac12)$, $m\in\N_0$ is the
usual harmonic oscillator spectrum.

We can now specify the generalized eigenfunctions corresponding to the
imaginary eigenvalues. Again multiplying $\h$ and $\ho$ with
$E'\,^2=(E/2)^2$ and working in the eigenbasis of $\vech{q}$ we have
\begin{eqnarray}
 \bra{q}E'\,^2\,
\ho\ket{n}=E'\,\big(n+\mbox{$\frac12$}\big)\,\psi_n^{\rm osc}(q) \ .
\end{eqnarray}
The orthonormal eigenfunctions of the harmonic oscillator Hamiltonian
are given by 
\begin{eqnarray}
 \psi_n^{\rm osc}(q)=N_n~\e^{-(E'/2)\,q^2}\,H_n\big(\sqrt{E'}~q\big) \ ,
\label{psinho}\end{eqnarray}
where $N_n=\big({\sqrt{E'}}\big/{2^n\,n!\,\sqrt{\pi}}\,\big)^{1/2}$ and
$H_n$ are the usual Hermite polynomials. Applying the operators
$\v{\pm}$ to these functions we get
\begin{eqnarray}
 f_n^\pm(q)\=\bra{q}\v{\mp\,\pi/4\theta}\ket{n}\=
\e^{\pm\,\frac{\i\pi}{8}}\,
\exp\big(\pm\,\mbox{$\frac{\i\pi}{4}$}\,q\,\p{q}\big)
\psi_n^{\rm osc}(q)\=
\e^{\pm\,\frac{\i\pi}{8}}\,\psi_n^{\rm
  osc}\big(\e^{\pm\,\frac{\i\pi}{4}}\,q\big) \ ,
\end{eqnarray}
and thus
\begin{eqnarray}
 f_n^\pm(q)&=&N_n^\pm~\e^{\mp\i(E'/2)\,q^2}\,H_n\big(\sqrt{\pm\i
   E'}~q\big)
\label{fnpmfinal}\end{eqnarray}
with $N_n^\pm=(\pm\i)^{1/4}\,N_n$. These functions belong to the
dual Schwartz space $\mathcal{S}'(\R)$.

We now note an important property. Since
$\v{\pm}^{-1}=\v{\mp}=\v{\mp}^\dagger$ and
$\ket{f_n^\pm}^\dagger=\bra{f_n^\pm}$ we have
\begin{eqnarray}
 \bra{f_n^\pm}\h\=\bra{n}\v{\pm}\,\h\=\bra{n}\v{\mp}^{-1}\,\h\=\mp\,
\eps_n\,\bra{f_n^\pm} \ , \label{nerd}
\end{eqnarray}
with $\eps_n :=(2\i/ E)\,(n+\frac12)$. Thus $\bra{f_n^\pm}$ is an
eigenbra of $\h$ corresponding to the generalized eigenvalue
$\mp\,\eps_n$. We will see that this subtle issue has some remarkable
consequences and will follow us through our entire treatment. Because
of this property, along with the orthonormality and completeness of
the eigenstates $\ket{n}$, we have
\begin{eqnarray}
 \bk{f_n^\pm}{f_m^\mp}\=\delta_{nm} \qquad \mbox{and} \qquad
\sum_{n=0}^\infty\,f_n^\pm(q)^*\,f_n^\mp(q'\,)\=\delta(q-q'\,) \ .
\end{eqnarray}

To further approach our goal of a discrete expansion we will
analytically continue the energy eigenfunctions $\chi_\pm^\eps$ and
$\eta_\pm^\eps$ into the complex energy plane and investigate their
analytic behaviours as functions of $\eps$. The distributions
$f_n^\pm$ will arise as residues of the functions $\chi^\eps_\pm$ and
$\eta_\pm^\eps$. We begin with the following lemma proven
in~\cite{chr04}.
\begin{lemma}
The parabolic cylinder functions $D_\lambda(z)$ are analytic functions
of $\lambda\in\C$.
\end{lemma}
The analytic structure of the functions \eqref{eta} is thus entirely
governed by the gamma-functions. Since the only singularities of
$\Gamma(\lambda)$ are simple poles at $\lambda=-n$, $n\in\N_0$ with
residues
\begin{eqnarray}
 \text{Res}_{\lambda=-n}\big(\Gamma(\lambda)\big)=\frac{(-1)^n}{n!} \ ,
\end{eqnarray}
and $\eps=\i E'\,(\nu+\frac12)$, we see that $\chi^\eps_\pm$ and
$\eta^\eps_\pm$ have poles at $\eps=-\i E'\,(n+\frac12)$ and $\eps=\i
E'\,(n+\frac12)$ with residues
\begin{eqnarray}
 \text{Res}_{\eps=-\i
   E'\,(n+\frac12)}\big(\chi_\pm^\eps(q)\big)&=&\frac{C}
{\sqrt{2\pi \,E'}}\,\frac{(-1)^n}{n!}\,{\i}^{-\frac n2-\frac{1}{4}}\,
D_n\big(\mp\,\sqrt{-2\i E'}~q\big) \ , \nonumber \\[4pt]
\text{Res}_{\eps=\i
   E'\,(n+\frac12)}\big(\eta_\pm^\eps(q)\big)&=&\frac{C}{\sqrt{2\pi\,
     E'}}\,\frac{(-1)^n}{n!}\,{\i}^{\frac n2+\frac{1}{4}}\,
D_{n}\big(\mp\,\sqrt{2\i E'}~q\big) \ .
\label{ResDn}\end{eqnarray}
Now using 
\begin{eqnarray}
 D_n(z)=2^{-n/2}~\e^{-z^2/4}\,H_n\big(z\big/\sqrt{2}\,\big)
\label{Dnz}\end{eqnarray}
for $n\in\N_0$, we find
\begin{eqnarray}
 \text{Res}_{\eps=-\eps_n}\big(\chi_\pm^\eps(q)\big)\=c_n^-\,f_n^-
 \qquad \mbox{and} \qquad
\text{Res}_{\eps=\eps_n}\big(\eta_\pm^\eps(q)\big)\=c_n^+\,f_n^+ \ ,
\end{eqnarray}
where the constants $c_n^\pm$ can be gleamed off from
(\ref{fnpmfinal}), (\ref{ResDn}) and (\ref{Dnz}).

We would now like to extend the integration over $\R$ to a closed
contour integral in \eqref{exp2}, and then apply the residue theorem
to get a discrete expansion. However, the integral over the arc at
infinity must not contribute to the contour integral. To characterize
this property, we introduce two Hardy classes of functions ${H}^2_\pm$
which may be defined as follows~\cite{gad04}. Given a function
$f(\eps)$ of the real variable $\eps$ which admits an analytic
continuation into the open upper complex half-plane, define the
function
\begin{eqnarray}
 I^+(y)=\int_{\R}\,\dd x~\big|f(x+\i y)\big|^2 \label{hardy1}
\end{eqnarray}
of $y>0$. Then $f(\eps)$ is in the Hardy class from
above ${H}^2_+$ if and only if the integrals \eqref{hardy1}
are uniformly bounded, or equivalently
\begin{eqnarray}
 \sup_{y>0}\,I^+(y)<\infty \ .
\end{eqnarray}
The Hardy class from below ${H}^2_-$ is defined in a similar
manner, by substituting $I^+(y)$ with the function
$I^-(y):=I^+(-y)$.

To make sense of the contour integral we define the spaces
\begin{eqnarray}
\Phi_-&=&\big\{\phi\in\mathcal{S}(\R_q)~\big|~
\bk{\chi_\pm^\eps}{\phi}\in\scal(\R_\eps)\cap{H}^2_-\big\} \ ,
\nonumber\\[4pt] 
\Phi_+&=&\big\{\phi\in\mathcal{S}(\R_q)~\big|~
\bk{\eta^{\eps}_\pm}{\phi}\in\scal(\R_\eps)\cap{H}^2_+\big\} \ ,
\label{Phipmdef}\end{eqnarray}
which are both dense in $L^2(\R)$. Using the residue theorem one then
proves the following result~\cite{chr03,chr04}.
\begin{theorem}
For any functions $\phi^\pm\in\Phi_\pm$, one has the expansions
\begin{eqnarray}
 \phi^\pm(q)=\sum_{n=0}^\infty\, \bk{f_n^\mp}{\phi^\pm}\,f_n^\pm(q)
 \ . 
\end{eqnarray}
\label{resonancethm}\end{theorem}
With these expansions we are now almost able to complete the proof of
the duality covariance of the quantum field theory in Minkowski
spacetime. What remains to show is how these expansions can be applied
to Schwartz functions in $\mathcal{S}(\R^2)$, such that each term
which arises is a generalized eigenfunction of the operators $\d^2$
and $\dt^2$. 

\subsection{Resonance expansion of Wigner
  distributions}\label{resexp2}

In order to achieve a discrete generalized eigenfunction expansion for
functions in an appropriate dense subspace of $L^2(\R^2)$, we will
again use the Weyl-Wigner correspondence. First of all, we have to
relate the domain of $\d^2$ and $\dt^2$  to the domain of $\h$. For
this, we define the space
\begin{eqnarray}
L^2(\R)\otimes L^2(\R)^\vee=\big\{\,\mbox{$\sum\limits_{k,l\in\N_0}$}\,
\kb{\psi_k}{\varphi_l}~\big|~\ket{\psi_k}\in
L^2(\R)\,,\,\bra{\varphi_l}\in L^2(\R)^\vee\,\big\} \ ,
\end{eqnarray}
which contains all possible linear combinations of tensor products
between functions in $L^2(\R)$ and its dual vector space
$L^2(\R)^\vee$. This space is isomorphic to $L^2(\R^2)$ and we may
switch between these spaces via the Weyl-Wigner correspondence. We may
thus identify $L^2(\R^2)$ with the space of Wigner distributions
$\{\,\wi{\vech{\phi}\,}~|~\vech{\phi}\in L^2(\R)\otimes
L^2(\R)^\vee\,\}$. In a similar vein, by restricting to compact
operators, we may identify the Schwartz space $\scal(\R^2)$ with
$\scal(\R)\otimes\scal(\R)^\vee$.
\begin{remark}
The integral representation~\cite{Gro46}
\beq
\wi{\,\kb{\psi}{\varphi}\,}=\frac1{2\pi}\,
\int_{\R}\, \dd k~ \e^{\i k\,x}\,\bk{t-\theta
   \,k/2}{\psi}\,\bk{\varphi}{t+\theta\, k/2}
\label{Wigenfn}\eeq
can be used to define the Wigner distribution of generalized
functions. In particular, it can be extended to a
map on the space of tempered distributions
${\sf W}:\mathcal{S}'(\R)\otimes\mathcal{S}'(\R)^\vee\to
\scal'(\R^2)$. Via (\ref{wewistar}), these extensions also define a
continuous star product on algebras of generalized functions.
\label{Wigenfnrem}\end{remark}

\begin{lemma}
The distributions $f_{n,m}^\pm(\vec{x})$ defined by
\begin{eqnarray}
f_{n,m}^\pm~:=~\wi{\,\ket{f_n^\pm}\bra{f_m^\mp}\,}\=
\wi{\,\v{\pm}\ket{n}\bra{m}\v{\pm}^{-1}\,}
\end{eqnarray}
are generalized eigenfunctions of $\d^2$ and $\dt^2$ with 
\begin{eqnarray}
 \d^2f_{n,m}^\pm(\vec{x})\=\pm\,\eps_n\, f_{n,m}^\pm(\vec{x}) \qquad 
 \mbox{and} \qquad
\dt^2f_{n,m}^\pm(\vec{x})&=&\pm\,\eps_m\, f_{n,m}^\pm(\vec{x}) \ ,
\label{ddteigeneqs}\end{eqnarray}
where
\beq
\eps_n=2\i E\,\big(n+\mbox{$\frac12$}\big) \ .
\eeq
\label{DtildeDeigenlemma}\end{lemma}
\begin{proof}
We use \eqref{dh} to find
\begin{eqnarray}
 \d^2f_{n,m}^\pm\=E^2\,\wi{\,\h\ket{f_n^\pm}\bra{f_m^\mp}\,}
 \qquad \mbox{and} \qquad
\dt^2f_{n,m}^\pm\=E^2\,\wi{\,\ket{f_n^\pm}\bra{f_m^\mp}\h\,} \ ,
\end{eqnarray}
and from \eqref{nerd} the generalized eigenvalue equations
(\ref{ddteigeneqs}) follow.
\end{proof}
\begin{remark}
One may wonder why we do not consider the more general functions
$f_{n,m}^{s,s'}$ given by
$f_{n,m}^{s,s'}=\wi{\v{s}\ket{n}\bra{m}\v{s'}^{-1}}$ with
$s,s'=\pm$. As is shown in Appendix~A
(Lemma~\ref{moregenlemma}), the distributions $f_{n,m}^{+,-}$ and
$f_{n,m}^{-,+}$ vanish identically, and only the generalized
eigenfunctions $f_{n,m}^\pm:=f_{n,m}^{\pm,\pm}$ remain.
\label{vanishrem}\end{remark}

The resonance expansion derived in Section~\ref{resexp1} above can now
be applied to Wigner distributions. For brevity, we will assume that
$\vech {\phi}\in \mathcal{S}(\R)\otimes \mathcal{S}(\R)^\vee$ is a
rank one operator $\vech{\phi}=\kb{\psi}{\varphi}$, but the extension
to general $\vech{\phi}$ follows straightforwardly by
linearity. Expanding $\vech{\phi}$ in parabolic cylinder functions, we
have either the expansion
\begin{eqnarray}
 \vech{\phi}&=&\sum_{s,s'=\pm}~\int_{\R}\,
 \dd\eps~\int_{\R}\,\dd\eps'~\ket{\chi_s^\eps}\,\bk{\chi_s^\eps}{\psi}\,
\bk{\varphi}{\chi_{s'}^{\eps'}}\,\bra{\chi_{s'}^{\eps'}}\label{chichi}
\end{eqnarray}
or
\begin{eqnarray}
 \vech{\phi}&=&\sum_{s,s'=\pm}~\int_{\R}\,
 \dd\eps~\int_{\R}\,\dd\eps'~\ket{\eta_s^\eps}\,\bk{\eta_s^\eps}{\psi}\,
\bk{\varphi}{\eta_{s'}^{\eps'}}\,\bra{\eta_{s'}^{\eps'}} \ . 
\label{etaeta}\end{eqnarray}
The other two possible combinations are excluded since they would
lead to expansions in the functions $f_{n,m}^{\pm,\mp}(\vec{x})$ for the
Wigner distributions, which vanish by Remark~\ref{vanishrem}
above. Now closing the contour of integration over $\eps$ in the lower
complex half-plane and over $\eps'$ in the upper complex half-plane in
the expansion \eqref{chichi}, and over $\eps$ in the upper half-plane
and over $\eps'$ in the lower half-plane in the expansion
\eqref{etaeta}, we find the resonance expansions
\begin{eqnarray}
 \vech{\phi}\=\sum_{n,m=0}^\infty\,\psi_n^+\,{\varphi_m^-}\,^*\,
\kb{f_n^-}{f_m^+} \qquad \mbox{and} \qquad
\vech{\phi}\=\sum_{n,m=0}^\infty\,\psi_n^-\,\varphi_m^+\,^*\,
\kb{f_n^+}{f_m^-}
\label{Weylresexp}\end{eqnarray}
on $\Phi_-\otimes\Phi_+^\vee$ and $\Phi_+\otimes\Phi_-^\vee$, respectively, 
where
\beq
\psi_n^\pm~:=~\bk{f_n^\pm}{\psi} \qquad \mbox{and} \qquad
{\varphi_m^\pm}\,^*~:=~\bk{\varphi}{f_m^\pm} \ .
\eeq
A detailed description of the appropriate domain for both expansions is
given in Section~\ref{time}.

\begin{theorem}
For the Wigner distributions
$\phi=\wi{\vech{\phi}\,}$, the resonance expansions correspond to
\begin{eqnarray}
 \phi(\vec{x})\=\sum_{n,m=0}^\infty\,\phi_{m,n}^+\,f_{m,n}^-(\vec{x}) \qquad
 \mbox{and} \qquad
\phi(\vec{x})\=\sum_{n,m=0}^\infty\,\phi_{m,n}^-\,f_{m,n}^+(\vec{x})
\label{Wignerresexp}\end{eqnarray}
on $\Phi_-\otimes\Phi_+^\vee$ and $\Phi_+\otimes\Phi_-^\vee$, respectively,
where
\begin{eqnarray}
\phi_{m,n}^\pm~:=~\bk{f_{m,n}^\pm}{\phi}\=
\int_{\R^2}\, \dd\vec{x}~f_{m,n}^\pm(\vec{x})^*\,\phi(\vec{x}) \ .
\end{eqnarray}
\label{Wignerresthm}\end{theorem}
\begin{proof}
On the one hand, using completeness we have
\begin{eqnarray}
 \phi(\vec{x})\=\sum_{n,m=0}^\infty\,
\wi{\,\ket{f_m^\mp}\,\bk{f_m^\pm}{\psi}\,\bk{\varphi}{f_n^\mp}\,
\bra{f_n^\pm}\,}(\vec{x})\=\sum_{n,m=0}^\infty\,
\psi_m^\pm\,{\varphi_n^\mp}\,^*\, f_{m,n}^\mp(\vec{x}) \ .
\end{eqnarray}
On the other hand, using (\ref{Wigenfn}) one has
\begin{eqnarray} 
f_{n,m}^\pm(\vec{x})^*\=\wi{\,\kb{f_n^\pm}{f_m^\mp}\,}(\vec{x})^*
\=\frac1{2\pi}\,
\int_{\R}\, \dd k~ \e^{\i k\,x}\,\bk{t-\theta
   \,k/2}{f_m^\mp}\,\bk{f_n^\pm}{t+\theta\, k/2}\=f_{m,n}^\mp(\vec{x}) \ ,
\nonumber \\
\label{Groid}\end{eqnarray}
and we get
\begin{eqnarray}
 \int_{\R^2} \,\dd\vec{x}~f_{m,n}^\pm(\vec{x})^*\,\phi(\vec{x})&=&
\int_{\R^2} \,\dd\vec{x}~f_{n,m}^\mp(\vec{x})\star\phi(\vec{x})
\nonumber\\[4pt] 
&=&\int_{\R^2}\, \dd\vec{x}~\sum_{k=0}^\infty\,
\wi{\,\ket{f_n^\mp}\,\bk{f_m^\pm}{\psi}\,\bk{\varphi}{f_k^\mp}\,
\bra{f_k^\pm}\,}(\vec{x}) \nonumber\\[4pt]&=&
\int_{\R^2} \,\dd\vec{x}~\sum_{k=0}^\infty\,
\psi_m^\pm\,{\varphi_k^\mp}\,^*\,f_{n,k}^\mp(\vec{x}) \ .
\end{eqnarray}
Since $\int_{\R^2}\, \dd\vec{x}~f_{k,l}^\pm(\vec{x})=\delta_{kl}$, the
result now follows.
\end{proof}
\begin{cor}
The resonance expansions in the space of Wigner distributions are
given by
\begin{eqnarray}
 \u\=\sum_{n,m=0}^\infty\,\kb{f_{n,m}^-}{f_{n,m}^+} \qquad \mbox{and}
 \qquad 
\u\=\sum_{n,m=0}^\infty\,\kb{f_{n,m}^+}{f_{n,m}^-}
\label{resexpsWig}\end{eqnarray}
on $\Phi_-\otimes\Phi_+^\vee$ and $\Phi_+\otimes\Phi_-^\vee$, respectively,
with the notation $f_{n,m}^\pm(\vec{x})=\bk{\vec{x}}{f_{n,m}^\pm}$.
\label{resexpcor}\end{cor}

\subsection{Regularization\label{reg}}

We are now ready to construct the duality invariant regularization of
our quantum field theory. As shown in Section~\ref{resexp2} above, instead
of a unique expansion as in Euclidean space~\cite{ls02}, we now have
two distinct resonance expansions (\ref{resexpsWig}) on the space of
Wigner distributions. However, it is easily checked using
Lemma~\ref{DtildeDeigenlemma} that both expansions individually lead
to a free action (\ref{freepart}) which is not manifestly real. We will
circumvent this problem in the following way, defering a detailed
technical analysis to Section~\ref{time}. The idea is to work on a
suitable dense domain $\Phi$ wherein \emph{both} resonance expansions
are possible. Naively, this space is the intersection of the spaces
$\Phi_-\otimes\Phi_+^\vee$ and $\Phi_+\otimes\Phi_-^\vee$, but this
definition is problematic due to the fact that the Hardy spaces have
trivial intersection $H_+^2\cap H_-^2=\{0\}$~\cite{gad04} (see
also~\cite[Prop.~4]{chr03}). In Section~\ref{Anprop} we will define
$\Phi$ more precisely.

The resonance expansion on the space $\Phi$ is given by inserting
\begin{eqnarray}
 \u=\frac12\,\sum_{s=\pm}~\sum_{n,m=0}^\infty\,\ket{f_{m,n}^s}
\bra{f_{m,n}^{-s}} \ .
\label{Phiresexp}\end{eqnarray}
One then readily checks the manifest reality of the action functional
$S_0[\phi]$ on the field domain $\Phi$ as 
\begin{eqnarray}
 S_0&=&\bra{\phi}\,\sigma\,\d^2+\sigmat\,\dt^2+
\mu^2\,\ket{\phi} \nonumber\\[4pt]
&=&\frac12\,\sum_{n,m=0}^\infty\,
\Big[\big(\sigma\,\eps_m+\sigmat\,\eps_n+
\mu^2\big)\,\bk{\phi}{f_{m,n}^+}\,
\bk{f_{m,n}^-}{\phi} \nonumber \\
&&\qquad\qquad\qquad +\,\big(-\sigma\,\eps_m-\sigmat\,\eps_n+
\mu^2\big)\,
\bk{\phi}{f_{m,n}^-}\,\bk{f_{m,n}^+}{\phi}\Big]
\label{realaction} \\[4pt]
&=&\frac12\,\sum_{n,m=0}^\infty\,\Big[\big(\sigma\,\eps_m+
\sigmat\,\eps_n+\mu^2\big)\,\phi_{m,n}^{+}{}^*\,
\phi_{m,n}^{-}+\big(\sigma\,\eps_m+\sigmat\,
\eps_n+\mu^2\big)^*\,
\phi_{m,n}^{-}{}^*\,\phi_{m,n}^{+}\Big] \ , \nonumber
\end{eqnarray}
where we have used ${\eps_n^*}=-\eps_n$. Thus both resonance
expansions together are required to yield a manifestly real action.  

In this basis the formal functional integration measure in
(\ref{ZJdef}) may be represented as
\beq
\mathcal{D}\phi~\mathcal{D}\phi^* = 
\prod_{n,m=0}^\infty~\prod_{s=\pm}\,\dd\phi_{n,m}^s~
\dd\phi_{n,m}^s{}^* \ ,
\label{intmeasrep}\eeq
and there are two non-vanishing free propagators given by
\begin{eqnarray}
C^\pm(n,m)\=\big\langle\,{\phi_{m,n}^\pm}^*\,
\phi_{m,n}^\mp\big
\rangle\=2\i\big(\pm\,(\sigma\,\eps_m+
\sigmat\,\eps_n)+\mu^2\big)^{-1} \ .
\label{Cpmnm1}\end{eqnarray}
They also arise by representing the operator
$(\sigma\,\d^2+\sigmat\,\dt^2+\mu^2)^{-1}$ in the two distinct basis
sets as 
\begin{eqnarray}
 C^\pm(n,m)=\bra{f_{m,n}^\mp}2\i\big(\sigma\,\d^2+
\sigmat\,\dt^2+\mu^2
\big)^{-1}\ket{f_{m,n}^\pm} \ .
\label{Cpmnm2}\end{eqnarray}
Thus we have two distinct propagators which, as we will see in
Section~\ref{time}, correspond to incoming and outgoing particle and
antiparticle asymptotic states. The spacetime representation of these
propagators, in the limiting case $\sigma=1$ of a background electric
field alone, is derived in Appendix~B in terms of confluent
hypergeometric functions.

Following~\cite{ls02}, the regularization scheme we shall employ is
based on the operator (\ref{D2Dt2}).
Each of the operators $\d^2$ and $\dt^2$ cut off the high energy modes
of one of the indices on the basis functions $f_{n,m}^\pm$. The
regulated propagators in Minkowski space are thus defined by
\begin{eqnarray}
 C_\Lambda^\pm(n,m)&:=&\bra{f_{m,n}^\mp}2\i\big(\sigma\,\d^2+
\sigmat\,\dt^2+\mu^2\big)^{-1}\,
L\big(\Lambda^{-2}\,|\d^2+\dt^2|\big)\ket{f_{m,n}^\pm}
\nonumber\\[4pt] &=&
\frac{2\i}{\pm\,\big(\sigma\,\eps_m+\sigmat\,\eps_n\big)+
\mu^2}\,L\left(\Lambda^{-2}\,
\left|\eps_n+\eps_m\right|\right) \ ,
\label{propreg}\end{eqnarray}
where $\Lambda\in\R$ is a cut-off parameter. The cut-off function $L$
is smooth and monotonically decreasing, with $L(y)=1$ for $y<1$ and
$L(y)=0$ for $y>2$. Since the differential operator
$\d^2+\dt^2$ is invariant under Fourier transformation up to a
rescaling, this regularization is duality invariant. 

In this basis each Feynman amplitude can be represented in the
schematic form
\begin{eqnarray}
 \sum_{n_1,m_1,\ldots,n_K,m_K=0}^\infty~\sum_{s_1,\dots,s_K=\pm}~
\prod_{k=1}^K\, C^{s_k}_\Lambda(n_k,m_k)\,(\cdots) \ ,
\end{eqnarray}
where $(\cdots)$ denotes the contributions from the noncommutative
interaction vertices derived from (\ref{intpart}) and combinatorial
factors. Since the propagator $C^{s}_\Lambda(n,m)$ given by
(\ref{propreg}) is nonzero only if
$|\eps_n+\eps_m|=2|E|\,(n+m+1)<2|\Lambda|$, which at finite $\Lambda$
is true solely for a finite number of distinct values of
$(n,m)\in\N_0^2$, every Feynman amplitude is represented by a finite
sum. This completes the proof of Theorem~\ref{thm2} and hence
establishes the quantum duality in Minkowski spacetime. 

\section{Configuration space}\label{time}

We will now clear up a few loose ends in our analysis of the previous
section. By Proposition~\ref{fnmexplprop}, the resonance expansion in
generalized eigenfunctions of the operators $\d^2$ and $\dt^2$ can be
obtained by just naively applying Wick rotations of the corresponding
results in Euclidean space. However, the expansion in Minkowski
signature involves a doubling of the effective field degrees of
freedom, which does not follow by a simple Wick rotation. We will
argue below that this doubling is due to a separation of time flow,
wherein one expansion corresponds to motion in a given time
direction while the other expansion corresponds to motion
in the opposite time direction. We will also give a precise definition
and rigorous, analytic description of the configuation 
space $\Phi$, and show that the restriction of the functional integral
to this domain may be regarded as an ingredient of the duality
invariant regularization, in the sense that $\Phi$ is a dense subspace
of $L^2(\R^2)$. 

\subsection{CT symmetry\label{Tempprop}}

We begin with a heuristic explanation for the doubling of degrees of
freedom ensuing from the resonance expansion (\ref{Phiresexp}) on the
configuration space $\Phi$. We recall that the two sets of
eigenfunctions in (\ref{eta}) are related by
$\eta_\pm^\eps(q)=\chi_\pm^\eps(q)^*$, so that the two subspaces in
(\ref{Phipmdef}) are related by $\Phi_+=\Phi_-{}^\dag$. It is for this
reason that each expansion in (\ref{resexpsWig}) on its own yields a
complex action, while the sum is manifestly real. On the other hand,
the transformation $\chi_\pm^\eps(q)\mapsto\eta_\pm^\eps(q)$ is
equivalent to the change $\nu+1\mapsto(\nu+1)^*=-\nu$ of the parameter
(\ref{nudef}). In turn, this is equivalent to reflection of the
electric field $E\mapsto-E$. Now using the explicit form of the
generalized eigenfunctions $f_{m,n}^\pm$
(Proposition~\ref{fnmexplprop}), we see that the time-reversal
operator $\vec{T}:t\mapsto-t$ leads to  
\begin{eqnarray}
 \vec{T}\,:\,f_{m,n}^\pm~\xrightarrow{t\mapsto-t}~f_{n,m}^\pm \ .
\end{eqnarray}
On the other hand, under the charge conjugation transformation
$\vec{C}:E\mapsto-E$ we get 
\begin{eqnarray}
\vec{C}\,:\, f_{m,n}^\pm(t,x)~\xrightarrow{E\mapsto-E}~
f_{m,n}^\mp(-t,x)=f_{n,m}^\mp(t,x) \ .
\end{eqnarray}
Thus by applying time-reversal plus charge conjugation we get the
mapping $f_{m,n}^\pm\mapsto f_{m,n}^\mp$. In particular, the spaces
$\Phi_+$ and $\Phi_-$ are in this way related via a
$\vec{C\,T}$-transformation, and one has
$\eta_\pm^\eps(q)=\vec{C\,T}\chi_\pm^\eps(q)$. 

The domain $\Phi$ is thus the smallest domain of fields
in which a $\vec{C\,T}$-invariant resonance expansion is possible. The
expansion coefficients in (\ref{Wignerresexp}) are related to each
other by
\beq
\phi_{n,m}^\mp\=\vec{C\,T}\phi_{n,m}^\pm~:=~
\bk{\vec{C\,T}f_{n,m}^\pm}{\phi} \ .
\label{phinmpmTrel}\eeq
After fixing a time orientation, we can thus interpret $C^+$ as the
propagator for incoming particles and $C^-$ as the propagator for
outgoing antiparticles. This is consistent with the properties
\beq
C^\pm(-\vec{x};-\vec{x}'\,)\=C^\pm(\vec{x};\vec{x}'\,) \qquad
\mbox{and} \qquad
C^\pm(t,x;t',x'\,)^*\=C^\mp(-t,x;-t',x'\,)
\label{Cpmprops}\eeq
which can be read off from (\ref{propspacetime}). Note that, by
Proposition~\ref{fnmexplprop}, the generalized eigenfunctions
$f_{m,n}^\pm$ have the asymptotic behaviour
\begin{eqnarray}
 f_{m,n}^\pm(t\rightarrow\pm\,\infty,x)\sim \e^{\mp\i
   E\,t^2}\times\text{ (polynomial in $t$) } \ .
\end{eqnarray}
This behaviour looks somewhat like the condition for outgoing and
incoming scattering states, except for the $t^2$ dependence in the
exponential and the polynomial factor which reflect the dipole nature
of the quanta in this field theory.

\subsection{Definition using Gel'fand-Shilov spaces\label{Anprop}}

We will now construct a suitable configuration space of fields
$\Phi\subset\scal(\R^2)$ which defines a rigged Hilbert space 
\beq
\Phi~\subset~L^2\big(\R^2\big)~\subset~\Phi' \ .
\label{PhiRHS}\eeq
This field domain will also define the space of matrices $\mcal$
to be integrated over in the matrix model of
Section~\ref{2matrixmodel}. As we demonstrate below, the appropriate 
configuration space $\Phi$ can be identified with a subalgebra of one
of the Gel'fand-Shilov spaces $\scal^\alpha_\alpha(\R^2)$ with
$\alpha\geq\frac12$, which are subspaces of Schwartz space
$\scal(\R^2)=\scal^\infty_\infty(\R^2)$. Their suitability rests on the
fact that they are closed under Fourier transformation and the
noncommutative star product, and their elements admit an expansion in
terms of the generalized eigenfunctions that we have constructed in
this paper.

We begin by reviewing the general definition of the Gel'fand-Shilov
spaces $\scal^\alpha_\alpha(\R^d)$, $d\geq1$~\cite{gel64}, and the
properties of them that we will need. This space is the set of all
smooth functions $\phi(\vec{q})$ on $\R^d$ for which there exists
constants $C>0$ and $M>0$ such that
\beq
\big\|\vec{q}^{\vec{m}}\,\partial_{\vec{q}}^{\vec{n}}
\phi\big\|_\infty\leq
C\,M^{|\vec{n}|+|\vec{m}|}\,\vec{n}^{\alpha\,\vec{n}}\,
\vec{m}^{\alpha\,\vec{m}}
\label{GSspacedef}\eeq
for all $\vec{n},\vec{m}\in\N_0^d$, where the norm is the usual
supremum norm on $L^\infty(\R^d)$. Here we use the conventional
multi-index notation where, for $\vec{n}=(n_1,\dots,n_d)\in\N_0^d$ and 
$\vec{q}=(q_1,\dots,q_d)\in\R^d$, we set
$\partial_{\vec{q}}^{\vec{n}}\phi(\vec{q})= \partial_{q_1}^{n_1}\cdots 
\partial_{q_d}^{n_d}\phi(\vec{q})$, $|\vec{n}|=n_1+\cdots+n_d$,
$\vec{n}^{\alpha\,\vec{n}}=n_1^{\alpha\,n_1}\cdots n_d^{\alpha\,n_d}$,
and so on (with the convention $n_i^{\alpha\,n_i}:=1$ for
$n_i=0$). The space $\scal^\alpha_\alpha(\R^d)$ can be realized as the
inductive limit of the family of Banach spaces
$\scal_\alpha^{\alpha,M}(\R^d)$, $M>0$ consisting of smooth functions
$\phi(\vec{q})$ on $\R^d$ with finite norm
\beq
\|\phi\|_{\alpha,M}:=
\sup_{\vec{n},\vec{m}\in\N_0^d}\,\frac{M^{|\vec{n}|+|\vec{m}|}}
{\vec{n}^{\alpha\,\vec{n}}\,\vec{m}^{\alpha\,\vec{m}}}\,
\big\|\vec{q}^{\vec{m}}\,\partial_{\vec{q}}^{\vec{n}}
\phi\big\|_\infty \ .
\label{Banachnorm}\eeq
The topology on $\scal^\alpha_\alpha(\R^d)$ is then the inductive
limit topology. This makes $\scal^\alpha_\alpha(\R^d)$ into a
Fr\'echet space which is a subspace of Schwartz space $\scal(\R^d)$.

The Fourier transform on $\scal^\alpha_\alpha(\R^d)$ is defined
analogously to (\ref{Fourier}), and it defines a topological
isomorphism. Thus the spaces $\GS(\R^d)$ form a family of Fourier
transform invariant spaces contained in the Schwartz class
$\scal(\R^d)$, which are closed under differentiation and
multiplication by a polynomial. They are thus well-suited as
configuration spaces for (free) duality covariant field theories. The
Gel'fand-Shilov spaces contain quasi-analytic classes, in the sense
that $\GS(\R^d)$ for $\frac12\leq\alpha\leq1$ are subspaces of the
space of entire functions on $\C^d$ restricted to $\R^d$. The smallest
non-trivial Gel'fand-Shilov space is $\scal^{1/2}_{1/2}(\R^d)$, which
contains, for example, the Gaussian fields
$\phi(\vec{q})=\e^{-\vec{q}^2}$. The spaces $\GS(\R^d)$ have been
previously proposed as suitable test function spaces for non-local
relativistic quantum field theories~\cite{bn04,Sol07a}.

The (strong) dual $\scal^\alpha_\alpha(\R^d)'$ of the Gel'fand-Shilov
class $\scal^\alpha_\alpha(\R^d)$ is a space of tempered
ultra-distributions of Roumieu type. It contains the space of tempered
distributions $\scal'(\R^d)$. The Fourier transform is extended to a
continuous linear transform on $\scal^\alpha_\alpha(\R^d)'$ by means
of the duality formula
\beq
\bk{\phi}{\mathcal{F}[F]}:=\bk{\mathcal{F}[\phi]}{F}
\label{Fourierdual}\eeq
for $F\in\scal^\alpha_\alpha(\R^d)'$ and
$\phi\in\scal^\alpha_\alpha(\R^d)$. It yields an isomorphism
$\scal^\alpha_\alpha(\R^d)'\to\scal^\alpha_\alpha(\R^d)'$.

Let us now specialize to the one-dimensional case $d=1$. Then the
topological algebras $\GS(\R)$ have the remarkable feature that the
harmonic oscillator eigenfunctions (\ref{psinho}) form a basis for the
expansion of fields in $\scal^\alpha_\alpha(\R)$~\cite{L-CP07}. Since
these eigenfunctions also form a complete orthonormal system in
$L^2(\R)$, it follows that the triplet of spaces
\beq
\GS(\R)~\subset~L^2(\R)~\subset~\GS(\R)'
\label{GSRHS}\eeq
is a rigged Hilbert space. The corresponding expansion coefficents
$\bk{n}{\phi}$ for $n\in\N_0$ and $\phi\in\GS(\R)$ may be
characterized as follows. The nuclear space $\mcal^\alpha_\alpha$ of
sequences of ultrafast falloff is the inductive limit of the family of
spaces $\mcal_\alpha^{\alpha,\kappa}$, ${\kappa>0}$ consisting of
complex sequences $\{a_n\}_{n\in\N_0}$ of finite norm
\beq
\big\|\{a_n\}\big\|_\kappa=\Big(\,\sum_{n=0}^\infty\,
|a_n|^2~\e^{2\Omega(\kappa\,\sqrt{n}\,)}\,
\Big)^{1/2} \ ,
\label{normkappa}\eeq
where we have defined the function
\beq
\Omega(y):=\sup_{n\in\N_0}\,\log\big(y^n\,n^{-\alpha\,n}\big)
\label{Omegay}\eeq
for $y>0$. That this space can be identified with the Gel'fand-Shilov
space $\GS(\R)$ is the content of the following crucial result, proven
in~\cite{L-CP07}.
\begin{theorem}
The mapping $\phi\mapsto a_n=\bk{n}{\phi}$, $n\in\N_0$ defines a
topological isomorphism on the spaces $\GS(\R)\to\mcal^\alpha_\alpha$.
\label{GSthm}\end{theorem}

When $a_n=\bk{n}{\phi}$ for $\phi\in\GS(\R)$, we will denote the norm
(\ref{normkappa}) by $\|\phi\|_\kappa$. This characterization leads to
the following result governing the generalized eigenfunction
expansions (\ref{exp2}), which enables us to replace \emph{both}
spaces $\Phi_\pm$ in (\ref{Phipmdef}) with the Gel'fand-Shilov space
$\GS(\R)$.
\begin{theorem}
For any function $\phi\in\GS(\R)$, one has:
\begin{itemize}
\item[(a)]
  $\displaystyle{\lim_{\eps\to\infty}\,\bkl{\eta^\eps_\pm}{\phi}=0}$, 
    where the limit is taken over generalized eigenvalues $\eps$ in
    the upper complex half-plane; and
\item[(b)] 
$\displaystyle{\lim_{\eps\to\infty}\,\bkl{\chi^\eps_\pm}{\phi}=0}$, 
    where the limit is taken over generalized eigenvalues $\eps$ in
    the lower complex half-plane.
\end{itemize}
\label{GSresexpthm}\end{theorem}
\begin{proof}
Since $\eta^\eps_\pm\in\scal'(\R)\subset\GS(\R)'$ and
$\phi\in\GS(\R)$, we have the Parseval equation~\cite{L-CP07}
\beq
\bkl{\eta^\eps_\pm}{\phi}=\sum_{n=0}^\infty\,
\bkl{\eta^\eps_\pm}{n}\,\bkl{n}{\phi}
\label{Parsevaleq}\eeq
with $\bk{\eta^\eps_\pm}{n}=\int_\R\,\dd
q~\eta^\eps_\pm(q)^*\,\psi_n^{\rm osc}(q)$. Using the Schwarz
inequality and Theorem~\ref{GSthm}, it follows that for every
$\kappa>0$ one has
\bea
\big|\bkl{\eta^\eps_\pm}{\phi}\big|&\leq&\sum_{n=0}^\infty\,
\big|\bkl{\eta^\eps_\pm}{n}\big|\,\big|\bkl{n}{\phi}\big|
\nonumber\\[4pt] &\leq&\Big(\,\sum_{n=0}^\infty\,
\big|\bkl{\eta^\eps_\pm}{n}\big|^2~\e^{-2\Omega(\kappa\,
\sqrt{n}\,)}\,\Big)^{1/2} \,\Big(\,\sum_{n=0}^\infty\,
\big|\bkl{n}{\phi}\big|^2~\e^{2\Omega(\kappa\,
\sqrt{n}\,)}\,\Big)^{1/2} \nonumber\\[4pt]
&=& \|\phi\|_\kappa\,\Big(\,\sum_{n=0}^\infty\,
\big|\bkl{\eta^\eps_\pm}{n}\big|^2~\e^{-2\Omega(\kappa\,
\sqrt{n}\,)}\,\Big)^{1/2} \ .
\label{etabd1}\eea
We will now substitute the explicit form of the generalized
eigenfunctions $\eta_\pm^\eps(q)$ from (\ref{eta}).

Using the integral representation (\ref{integral_representation}) for
the parabolic cylinder functions, it is straightforward to derive the
integral identity
\beq
\int_\R\,\dd t~D_\nu(t)=-\frac{2\,\sqrt{\pi}~2^{\frac12\,(\nu+1)}}
{\nu\,\Gamma\big(-\frac12\,\nu\big)} \ .
\label{parcylintid}\eeq
We will also use the estimate
\beq
\big\|\psi_n^{\rm osc}\big\|_\infty\leq C\,n^k \ ,
\label{hoest}\eeq
for some constants $C>0$ and $k\in\N$ which are independent of
$n$. For brevity, in what follows we use the same symbol $C$ to absorb
all constants independent of $n$ and of the complex parameter $\nu$ in
(\ref{nudef}). We then find the bound
\beq
\big|\bkl{\eta^\eps_\pm}{n}\big|\leq C\,n^k\,\Big|\,
\frac{\e^{-\i\pi\,\nu/4}\,2^{\nu/2}\,\Gamma(-\nu)}
{\nu\,\Gamma\big(-\frac12\,\nu\big)}\,\Big| \ .
\label{etanbd1}\eeq

Using the Stirling expansion of the gamma-functions for
$|\nu|\to\infty$ and the definition (\ref{Omegay}), we then have
\bea
\big|\bkl{\eta^\eps_\pm}{n}\big|^2~\e^{-2\Omega(\kappa\,
\sqrt{n}\,)}&\leq& C\,n^{2k}\,\big|\e^{\i\pi\,\nu/2}\,\nu^{-\nu-2}~
\e^\nu\big|~\e^{-2\Omega(\kappa\,\sqrt{n}\,)} \nonumber\\[4pt]
&=&C\,\big|\e^{\i\pi\,\nu/2}\,\nu^{-\nu-2}~\e^\nu\big|\,
\frac{n^{2k}}{\Big(\,\sup\limits_{m\in\N_0}\,
\kappa^m\,n^{m/2}\,m^{-\alpha\,m}\,\Big)^2} \nonumber\\[4pt]
&\leq&C\,\big|\e^{\i\pi\,\nu/2}\,\nu^{-\nu-2}~\e^\nu\big|\,
\frac{n^{2k}\,(2k+2)^{4\alpha\,(k+1)}}{\kappa^{4(k+1)}\,
n^{2k+2}} \nonumber\\[4pt] &\leq&
C\,\big|\e^{\i\pi\,\nu/2}\,\nu^{-\nu-2}~\e^\nu\big|\,
\frac1{n^2} \ ,
\label{etaOmbd1}\eea
where we have chosen $\kappa\geq(2k+2)^\alpha$. Substituting
(\ref{etaOmbd1}) back into (\ref{etabd1}), since the series
$\sum_{n\in\N}\,\frac1{n^2}$ converges we have finally
\beq
\big|\bkl{\eta^\eps_\pm}{\phi}\big|\leq C\,\|\phi\|_\kappa\,
\big|\e^{\i\pi\,\nu/4}\,\nu^{-\frac12\,\nu-1}~\e^{\nu/2}\big| \ .
\label{etabdfinal}\eeq
The right-hand side of (\ref{etabdfinal}) vanishes in the limit ${\rm
  Re}(\nu)\to+\infty$, which proves (a). With the same techniques, an
analogous bound for $|\bk{\chi^\eps_\pm}{\phi}|$ is obtained using
the explicit form for the generalized eigenfunctions
$\chi_\pm^\eps(q)$ in (\ref{eta}), which now vanishes for ${\rm
  Re}(\nu)\to-\infty$ and establishes (b).
\end{proof}
\begin{cor}
For any function $\phi\in\GS(\R)$, one has the resonance expansion
\beq
\phi(q)=\frac12\,\sum_{s=\pm}~\sum_{n=0}^\infty\,
\bkl{f_n^{-s}}{\phi}\,f_n^s(q) \ .
\label{GSresexp}\eeq
\label{GSresexpcor}\end{cor}
\begin{proof}
From the estimate (\ref{etabdfinal}) and the analogous one for
$|\bk{\chi^\eps_\pm}{\phi}|$, together with the explicit forms of the
generalized eigenfunctions in (\ref{eta}), we see that the integrands
of (\ref{exp2}) evaluated on an arc of radius $r\to\infty$ in the
upper or lower half-planes respectively vanish
much faster than $r^{-1-\epsilon}$ for $\epsilon>0$. As in the proof
of~\cite[Thm.~2]{chr03}, the contributions to the contour
integrals from the arcs at infinity thus vanish.
\end{proof}

We can now transport the resonance expansion (\ref{GSresexp}) to the
appropriate space of Wigner distributions, exactly as we did in
Section~\ref{resexp2}. The following result, whose proof may be found
in~\cite{teo06}, is helpful for this purpose.
\begin{lemma}
Let $\psi,\varphi\in\GS(\R)'$. Then $\psi\in\GS(\R)$ if and only if
$\wi{\,\kb{\psi}{\varphi}\,}\in\GS(\R^2)$.
\label{WigGSlemma}\end{lemma}

It follows from Lemma~\ref{WigGSlemma} that the Wigner distribution
(\ref{Groid}) induces a transformation
\beq
{\sf W}\,:\,\GS(\R)\otimes\GS(\R)'~\longrightarrow~\GS\big(\R^2\big)
\label{WigGStransf}\eeq
for $\alpha\geq\frac12$. In this way the space of duality covariant
noncommutative scalar fields can be identified with the
subspace $\Phi={\sf W}\big(\GS(\R)\otimes\GS(\R)^\vee\,\big)$ of the
Gel'fand-Shilov space $\scal^\alpha_\alpha(\R^2)$. This defines a
topological algebra which is continuously closed under the star
product, because of the projector property $f_{n,m}^\pm\star
f_{k,l}^\pm=\delta_{mk}\,f_{n,l}^\pm$. This property is consistent
with the result of~\cite{Sol07a} that the star product has a unique
continuous extension to any Gel'fand-Shilov space
$\scal^\alpha_\alpha(\R^2)$. Using Theorem~\ref{GSthm}, the
corresponding space of sequences $\{\phi_{n,m}^\pm\}_{n,m\in\N_0}$ can
be identified with a subspace $\mcal$ of the nuclear space of
sequences on $\N_0^2$ of ultrafast falloff~\cite{L-CP07}. 

\section{The two-matrix model\label{2matrixmodel}}

As in the Euclidean case~\cite{lsz04}, the formalism developed in this
paper enables a reformulation of the duality covariant quantum field
theory as a succinct matrix model, though now with some crucial
differences. Thus far we have used the Weyl-Wigner correspondence at
the self-dual point $\theta=2/E$ to find the generalized eigenbasis of
the operator $\d^2$, which depends on the absolute value of the
electric field $E$. However, the star product appearing in the
interaction term (\ref{intpart}) does not depend on the electric
field, but on the noncommutativity parameter $\theta$. The generalized
eigenbasis of $\d^2$ has a very nice projector property under the star
product if the noncommutativity parameter is set equal to $2/E$. For
$\theta\neq2/E$, this is no longer true in general, while the free
part of the action (\ref{freepart}) is still diagonal. We will now
reverse the logic. We will suppose that the basis functions
$f_{n,m}^\pm$ are defined with respect to the electric field
$2/\theta\neq E$. In this case the complete action can be mapped onto
a coupled complex two-matrix model.

We will fix $E\neq\pm\,2/\theta$ generically, and assume that
$E,\theta>0$. The interaction part of our action (\ref{intpart}) can
be mapped onto a matrix model by noting the identities 
\begin{eqnarray}
 \int_{\R^2}\, \dd\vec{x}~f_{n,m}^s(\vec{x})&=&\delta_{nm} \ ,
 \nonumber\\[4pt] 
f_{n,m}^s\star f_{k,l}^{{s}}&=&\delta_{mk}\,f_{n,l}^s \ , \nonumber
\\[4pt]
f_{n,m}^s\,^*&=&f_{m,n}^{-s}
\end{eqnarray}
for $s=\pm$. Thus the only surviving combinations for star products of
four distributions $f_{n,m}^s$ are
\begin{eqnarray}
 f_{n_1,m_1}^\pm\star f_{n_2,m_2}^\pm\star f_{n_3,m_3}^\pm\star
 f_{n_4,m_4}^\pm=\delta_{n_2,m_1}\,\delta_{n_3,m_2}\,\delta_{n_4,m_3}~
f^\pm_{n_1,m_4} \ .
\end{eqnarray}
By using one of the two expansions
$\phi(\vec{x})=\sum_{n,m\in\N_0}\,f_{n,m}^s(\vec{x})\,\phi^{-s}_{n,m}$
for the scalar fields, we can express the interaction term
$\int_{\R^2}\, \dd\vec{x}~
(\phi^*\star\phi\star\phi^*\star\phi)(\vec{x})$ as a
matrix product
$\Tr\big(\phi^\dagger_{-s}\,\phi_s\,\phi_{-s}^\dagger\,\phi_s\big)$
for $s=\pm$ and $ (\phi_s)_{n,m}:=\phi_{n,m}^s$. The interaction
$\int_{\R^2}\,\dd\vec{x}~
(\phi^*\star\phi^*\star\phi\star\phi)(\vec{x})$ gives
$\Tr\big(\phi^\dagger_{-s}\,\phi^\dagger_{-s}\,
\phi_{{s}}\,\phi_s\,\big)$. On the domain $\Phi$,
the action (\ref{intpart}) can thus be written as a matrix model
\beq
S_{\rm int}=\frac12\,\sum_{s=\pm}\,\Tr\big(\alpha\,
\phi^\dagger_{-s}\,\phi_s\,\phi_{-s}^\dagger\,\phi_s
+\beta\,\phi^\dagger_{-s}\,\phi^\dagger_{-s}\,
\phi_{{s}}\,\phi_s\,\big) \ .
\label{intmatrix}\eeq

The free action (\ref{freepart}) for $E\neq\pm\,2/\theta$ can also be
written as a matrix product in the following way. With the help of
the operators \eqref{a2} we can write
\begin{eqnarray}
 \d^2=\mbox{$\frac{1}{4\theta}$}\,\left[(2+\theta\,E)^2\,\big
(a_1^+\,a_1^-+\mbox{$\frac{\i}{2}$}\big)+(2-\theta\,E)^2\,\big
(a_2^+\,a_2^-+\mbox{$\frac{\i}{2}$}\big)+\big(\theta^2\,E^2-4\big)\,
\big(a_1^+\,a_2^++a_1^-\,a_2^-\big)\right] \label{d} \ .
\end{eqnarray}
Note that at the two self-dual points $\theta=\pm\,2/E$ the free
action simplifies considerably, since
\begin{eqnarray}
\d^2&=&2E\,\big(a_1^+\,a_1^-+\mbox{$\frac\i2$}\big) \qquad \mbox{for}
\quad \theta\=2/E \ , \nonumber\\[4pt]
\d^2&=&2E\,\big(a_2^+\,a_2^-+\mbox{$\frac\i2$}\big) \qquad \mbox{for}
\quad \theta\=-2/E \ .
\end{eqnarray}
The corresponding expressions for the operator $\dt^2$ are obtained by
interchanging $a_1^\pm\leftrightarrow a_2^\pm$ above. Using
\eqref{state4} we find 
\begin{eqnarray}
 a_1^\pm\,a_2^\pm f_{n,m}^s&=&\i s\,\sqrt{n+\mbox{$\frac12\pm\frac s2$}}~
\sqrt{m+\mbox{$\frac12\pm\frac s2$}}~f_{n\pm s,m\pm s}^s  \ ,
\end{eqnarray}
with the abbreviated notation $n\pm s:=n\pm1$ for $s=+$ and 
$n\pm s:=n\mp1$ for $s=-$.
The free action of our model can thus be expressed as
\begin{eqnarray}
S_0&=& \frac{1}{8\theta}\,\sum_{s=\pm}\,\Tr\Big(4\theta\,\mu^2\,
\phi_s^\dag\,\phi_{-s}+\big(\theta^2\,E^2-4\big)\,
\big(\phi_s^\dagger\,\Gamma_{s}^\dagger\,
\phi_{-s}\,\Gamma_{s}+\phi_{-s}\,
\Gamma_{s}^\dagger\,\phi_s^\dagger\,\Gamma_{s}\big) 
\label{free2matrix}\\ && \qquad\qquad\qquad
+\,s\,\big((2-\theta\,E)^2+8\sigma\,\theta\,E\big)\,
\phi_s^\dagger\,\eps\,\phi_{-s}+s\,\big((2+\theta\,E)^2-8
\sigma\,\theta\,E\big)\,
\phi_{{-s}}\,\eps\,\phi^\dagger_{s}\Big) \ , \nonumber
\end{eqnarray}
where we have introduced the matrices
\begin{eqnarray}
(\Gamma_{s})_{n,m}\=\i s\,\sqrt{m+1}~
\delta_{n,m+1} \qquad \mbox{and} \qquad
(\eps)_{n,m}\=\i\big(n+\mbox{$\frac12$}\big)\,\delta_{n,m} \ .
\end{eqnarray}
This action has a similar structure to that of the Euclidean
case~\cite{lsz04}.

The domain of this matrix model is the space of (infinite) matrices
$\mcal$ described in Section~\ref{Anprop}. Let us now set $\alpha=1$,
$\beta=0$ in the interaction term (\ref{intmatrix}), $\sigma=1$ in the
free part (\ref{free2matrix}),
and consider the matrix model at the self-dual point
$\theta=+2/E$. The full action is then given by
\beq
S_{\vee}=\frac12\,\sum_{s=\pm}\,\Tr\Big(4s\,\theta^{-1}\,\phi_s^\dag\,
\eps\,\phi_{-s}+\mu^2\,\phi_s^\dag\,\phi_{-s}+g^2\,\big(
\phi_s^\dag\,\phi_{-s}\big)^2\Big) \ .
\label{actionselfdual}\eeq
The matrix model at the other self-dual point $\theta=-2/E$ 
is gotten by interchanging
$\phi_s^\dag\leftrightarrow\phi_{-s}$ in (\ref{actionselfdual}). The
action (\ref{actionselfdual}) admits a continuous $GL(\infty)\times
GL(\infty)$ symmetry group defined by the transformations 
\beq
\phi_s~\longmapsto~\phi_s\,U_s \qquad \mbox{and} \qquad
\phi_s^\dag~\longmapsto~U_{-s}^{-1}\,\phi_s^\dag \ ,
\label{GLsymm}\eeq
with $U_\pm\in\mcal\cap GL(\infty)$. Thus the self-dual matrix
model describes an integrable quantum field theory, just as in the
Euclidean case~\cite{lsz03,lsz04}. By~\cite{lsz01} and the discussion
of Section~\ref{Tempprop}, the unitary $U(\infty)\times U(\infty)$
subgroup of this symmetry group consists of matrices $U_\pm$
corresponding to canonical transformations of $\R^2$ along the
forward/backward light-cone direction. Note that the self-dual matrix
model is also invariant under a discrete $\Z_2$ symmetry group
generated by the combined time-reversal and charge conjugation 
transformation
\beq
\vec{C\,T}\,:\,\big(\phi_s\,,\,\phi_s^\dag\,\big)~\longmapsto~
\big(\phi_{-s}\,,\,\phi_{-s}^\dag\big) \quad , \quad
\theta~\longmapsto~-\theta \ .
\label{CTselfdual}\eeq 

\section{Generalization to higher dimensions\label{HigherD}}

There is a natural UV/IR-duality invariant extension of our
$1+1$-dimensional model to higher dimensional Minkowski spacetime,
which combines our result with that of~\cite{ls02} for the Euclidean
case. We will demonstrate this in $D=2d+2$ dimensional
Minkowski spacetime with coordinates $\vec{x}=(x^\mu)$,
$\mu=0,1,\ldots,2d+1$, $x^0=t$ and derivatives
$\p{\mu}=\partial/\partial x^\mu$. The extended field theory in
$D$-dimensional Minkowski spacetime has a similar form as before.

The interactions are formally the same as in
(\ref{intpart}), while the free part of the action now reads
\begin{eqnarray}
 S_0&=&\int_{\R^D}\, \dd\vec{x}~\phi^*(\vec{x})\big(\sigma\,\k^2+
\sigmat\,\kt^2+\mu^2\big)\phi(\vec{x})
\end{eqnarray}
with $\k^2:=\frac{1}{2}\,(-\i\p{\mu}+F_{\mu\nu}\,x^\nu)^2$ and the
$D\times D$ antisymmetric electromagnetic tensor in Jordan
normal form
\begin{eqnarray}
 (F_{\mu\nu})=\begin{pmatrix} 0 & E & & & & & \\
-E & 0 & & & & & \\
 & & 0 & B_1 & & & \\ & & -B_1 & 0 & & & \\
 & & & & \ddots& & \\
 & & & & &0 & B_d \\ & & & & &
-B_d & 0 \end{pmatrix}
\end{eqnarray}
for $E,B_k>0$. The differential operator $\kt^2$ is defined below. The
coordinate system on $\R^D$ is chosen in such a way that the 
noncommutativity parameter matrix $(\theta^{\mu\nu})$ appearing in the
star product is in its canonical skew-diagonal form
\begin{eqnarray}
\big(\theta^{\mu\nu}\big)= \begin{pmatrix} 0 & \theta_0 & & & & & \\
-\theta_0 & 0 & & & & &\\
& & 0 & \theta_1 & & & \\ & & 
-\theta_1 & 0 & & & \\ & &
& & \ddots & & \\
& & & & & 0 & \theta_d\\ & & & & & -\theta_d & 0 \end{pmatrix} \ ,
\end{eqnarray}
with $\theta_a>0$.

With these definitions the operator $\k^2$ decomposes into a sum
\begin{eqnarray}
 \k^2=\d^2+\sum_{k=1}^d\,\d^2_{{\rm E},k}
\end{eqnarray}
of $d$ copies of the {Landau Hamiltonian} 
\begin{eqnarray}
\d^2_{{\rm E},k}=\mbox{$\frac{1}{2}$}\,\left[-\big(\partial_{2k}^2+
\partial_{2k+1}^2\big)-2\i B_k\,\big(x^{2k+1}\,\partial_{2k}-x^{2k}\,
\partial_{2k+1}\big)+B_k^2\,\left(\big(x^{2k}\big)^2+
\big(x^{2k+1}\big)^2\right)\right] 
\end{eqnarray}
for $k=1,\ldots,d$, and the Klein-Gordon operator $\d^2$ introduced in
Section~\ref{Formulation}. The classical duality is now proven in
the same way as before. The self-dual point is given by
$(\theta^{\mu\nu})=\pm\,2(F_{\mu\nu})^{-1}$, or equivalently by
$\theta_0=\pm\,2/E$ and $\theta_k=\pm\,2/B_k$, $k=1,\dots,d$, where
the sign has to be the same for all $\theta_ a $, $ a =0,1,\ldots,d$.

Eigenfunctions of the operator $\k^2$ are now given by tensor products
of eigenfunctions of $\d^2$ and $\d^2_{{\rm E},k}$ for $k=1,\ldots,d$,
analogously to the Euclidean analysis of~\cite{ls02}. The Landau
Hamiltonian $\d^2_{{\rm E},k}$ describes the motion of a charged
particle in the two-dimensional Euclidean $(x^{2k},x^{2k+1})$-plane in
the presence of a background magnetic field with field strength
$2B_k$, and its eigenfunctions are the well-known Landau wavefunctions
$f_{m_k,n_k}(x^{2k},x^{2k+1})$, $m_k,n_k\in\N_0$. These functions are
simultaneous eigenfunctions of the operators $\d^2_{{\rm E},k}$ and
$\dt^2_{{\rm E},k}:=\d^2_{{\rm E},k}\big|_{B_k\to-B_k}$ with
eigenvalues $2B_k\,(m_k+\frac12)$ and $2B_k\,(n_k+\frac12)$,
respectively. These definitions give rise to a new operator $\kt^2$
obtained from $\k^2$ by substituting $\dt^2$ for $\d^2$ and
$\dt^2_{{\rm E},k}$ for $\d^2_{{\rm E},k}$. Simultaneous generalized
eigenfunctions of $\k^2$ and $\kt^2$ are therefore given by tensor
products
\begin{eqnarray}
f_{\vec{p}}^\pm(\vec{x})=\big(
f_{m_0,n_0}^\pm\otimes f_{m_1,n_1}\otimes\cdots\otimes f_{m_d,n_d}
\big)(\vec{x})\label{extlandau}
\end{eqnarray}
with $\vec{p}:=(\vec{m},\vec{n})=
(m_0,m_1,\ldots,m_d,n_0,n_1,\ldots,n_d)\in\N_0^D$. The
corresponding generalized eigenvalue equations are 
\begin{eqnarray}
 \k^2f_{\vec{p}}^\pm&=&\Big(\pm\,2\i E\,\left(m_0+\mbox{$\frac{1}{2}$}
\right)+\sum_{k=1}^d\,2B_k\,\left(m_k+\mbox{$\frac{1}{2}$}\right)
\Big)\,f_{\vec{p}}^\pm~=:~E^\pm(\vec{m})\,f_{\vec{p}}^\pm \ ,
\nonumber \\[4pt]
\kt^2f_{\vec{p}}^\pm&=&\Big(\pm\,2\i E\,\left(n_0+\mbox{$\frac{1}{2}$}
\right)+\sum_{k=1}^d\,2B_k\,\left(n_k+\mbox{$\frac{1}{2}$}\right)
\Big)\,f_{\vec{p}}^\pm~=:~E^\pm(\vec{n})\,f_{\vec{p}}^\pm \ .
\end{eqnarray}
This extended field theory now comprises all the features of our
$1+1$-dimensional model and the $2d$-dimensional Euclidean model
investigated in~\cite{ls02}. Thus it is duality covariant and
has a matrix model representation in terms of the \emph{extended
  Landau basis} defined in \eqref{extlandau}. 

The Landau wavefunctions $f_{m,n}(x,y)$ form a basis for $L^2(\R^2)$,
which simply reflects the fact that they are the Wigner distributions
of the harmonic oscillator eigenoperators $\kb{m}{n}$. The extended
Landau wavefunctions $f_{\vec{p}}^\pm$ are Wigner distributions of
the tensor products
\beq
\kb{f_{m_0}^\pm,m_1,\ldots,m_d}{f_{n_0}^\mp,n_1,\ldots,n_d}=
\kb{f_{m_0}^\pm}{f_{n_0}^\mp}\otimes\kb{m_1}{n_1}\otimes\cdots
\otimes\kb{m_d}{n_d} \ .
\label{Wigtensprod}\eeq
Most of
the analysis of the $1+1$-dimensional case is now easily generalized
to higher dimensions. Each field $\phi$ in a suitable domain
$\Phi\subset\scal(\R^D)$, dense in $L^2(\R^D)$, can be
decomposed as
\begin{eqnarray}
 \phi(\vec{x})&=&\frac12\,\sum_{s=\pm}~\sum_{\vec{p}\in\N_0^D}\,
f_{\vec{p}}^s(\vec{x})\,\phi_{\vec{p}}^{-{s}}
\qquad \mbox{with} \quad \phi_{\vec{p}}^s\=\int_{\R^D}\,
 \dd\vec{x}~f_{\vec{p}}^s(\vec{x})^*\, \phi(\vec{x}) \ .
\end{eqnarray}
The free action takes the form
\begin{eqnarray}
 S_0=\frac12\,\sum_{s=\pm}~\sum_{\vec{p}\in\N_0^D}\,\big(
\sigma\,E^s(\vec{m})+\sigmat\,E^s(\vec{n})+
\mu^2\big)\,\phi_{\vec{p}}^s\,^*\,\phi_{\vec{p}}^{-{s}} \ ,
\end{eqnarray}
and the two propagators in this basis, given by
\begin{eqnarray}
 C^\pm(\vec{p})&=&\bra{f_{\vec{p}}^\mp}2\i\big(
\sigma\,\k^2+\sigmat\,\kt^2+\mu^2\big)^{-1}
\ket{f_{\vec{p}}^\pm} \nonumber\\[4pt] &=& 2\i\big(
\sigma\,E^\pm(\vec{m})+\sigmat\,
E^\pm(\vec{n})+\mu^2\big)^{-1} \ ,
\end{eqnarray}
can be regularized in the same way as before, proving the duality
invariance of the higher-dimensional regularized quantum field
theory. 

Using the noncommutativity parameters $(\theta^{\mu\nu})$ for the
Wigner transformations and the star product, the resulting extended
Landau wavefunctions obey the same nice projector property under the
star product given by
\begin{eqnarray}
 f_{(\vec{m},\vec{n})}^\pm\star f^\pm_{(\vec{m}',\vec{n}'\,)}=
\delta_{\vec{n},\vec{m}'}\,f_{(\vec{m},\vec{n}'\,)}^\pm \ ,
\label{projhigher}\end{eqnarray}
where $\delta_{\vec{m},\vec{n}}:=
\delta_{m_0,n_0}\,\delta_{m_1,n_1}\cdots\delta_{m_d,n_d}$. As was
shown in \cite{lsz04}, the Landau Hamiltonians $\d^2_{{\rm E},k}$ and
$\dt^2_{{\rm E},k}$ can be written in terms of standard harmonic
oscillator creation and annihilation operators for each
$k=1,\dots,d$. With the definitions
\begin{eqnarray}
 (\eps_0^s)_{\vec{m},\vec{n}}&=&\i
 s\,\left(n_0+\mbox{$\frac{1}{2}$}\right)\,
\delta_{\vec{m},\vec{n}} \ , \nonumber\\[4pt]
(\eps_k^s)_{\vec{m},\vec{n}}&=&\left(n_k+\mbox{$\frac{1}{2}$}\right)\,
\delta_{\vec{m},\vec{n}} \ , \notag\\[4pt]
(\Gamma_0^s)_{\vec{m},\vec{n}}&=&\i
s\,\sqrt{n_0+1}~\delta_{m_0,n_0+1}\,
\delta_{m_1,n_1}\cdots\delta_{m_d,n_d} \ , \nonumber\\[4pt]
(\Gamma_k^s)_{\vec{m},\vec{n}}&=&\sqrt{n_k+1}~\delta_{m_0,n_0}
\cdots\delta_{m_k,n_k+1}\cdots\delta_{m_d,n_d}
\end{eqnarray}
along with $F_0=E$ and $F_k=B_k$ for $k=1,\ldots,d$, we can thereby
map the free part of the 
$D$-dimensional $\phi^{\star 4}$ field theory in
Minkowski spacetime onto a two-matrix model with action
\begin{eqnarray}
 S_0&=&\frac{1}{8}\,\sum_{s=\pm}~\sum_{ a =0}^d\,\frac1{\theta_a }\,
\Tr\Big(\,\frac{4\theta_a\,\mu^2}{d+1}\,\phi_s^\dagger\,\phi_{-{s}}+
\big(\theta_a^2\,F_ a ^2-4\big)\,\big(\phi^\dagger_s\,
\Gamma_ a ^{s}\,^\dagger\,\phi_{-{s}}\,\Gamma_ a ^s+\phi_{-{s}}\,
\Gamma_ a ^{s}\,^\dagger\,\phi^\dagger_s\,\Gamma_ a ^s\big) \\
&& \qquad \qquad 
+\,\big((2-\theta_a\, F_ a )^2+8\sigma\,\theta_a\,F_a\big)\,
\phi^\dagger_s\,\eps_ a ^s\,\phi_{-{s}}+
\big((2+\theta_a\, F_ a )^2-8\sigma\,\theta_a\,F_a\big)\,
\phi_{-{s}}\,\eps_ a ^s\,\phi^\dagger_s\Big) \ , \nonumber
\end{eqnarray}
where $(\phi_s)_{\vec{m},\vec{n}}:=\phi_{(\vec{m},\vec{n})}^s$
(regarded as a matrix via lexicographic ordering $\N_0^D\sim\N_0$, for
example) and
$\Tr(\phi_s):=\sum_{\vec{n}\in\N_0^D}\,(\phi_s)_{\vec{n},\vec{n}}$. 
Due to (\ref{projhigher}), the interaction terms take the same form as
in (\ref{intmatrix}). At the self-dual points given by $\theta_ a
=\pm\,2/F_ a$, and with the definition
\begin{eqnarray}
 (\eps^s)_{\vec{m},\vec{n}}&=&\sum_{ a =0}^d\,2F_ a \,(\eps_ a
 ^s)_{\vec{m},\vec{n}} \= 
2\Big(\i s\,E\,\left(n_0+\mbox{$\frac{1}{2}$}\right)+\sum_{k=1}^d\,
B_k\left(n_k+\mbox{$\frac{1}{2}$}\right)\Big)\,\delta_{\vec{m},\vec{n}}
\ ,
\end{eqnarray}
we obtain the same self-dual 
two-matrix model as in Section~\ref{2matrixmodel}.

\section{Summary and discussion\label{Conclusions}}

In this paper we have proven that a noncommutative
$\phi^{\star4}$-theory, describing a charged scalar boson moving in
Minkowski spacetime in the presence of a background electromagnetic
field, is invariant under a special UV/IR duality generated by
symplectic Fourier transformation of fields. This was achieved 
by extending the methods which were used in the Euclidean
situation~\cite{ls02}. We were able to map our noncommutative field
theory onto a two-matrix model and regularize it in a duality
invariant fashion. What makes the Minkowskian story much more
intricate than the Euclidean one is that the Lorentzian kinetic
operator has a continuous spectrum extended over the whole real
line. By analytically continuing its eigenfunctions into the complex
energy plane and closing the integration contour of the continuous
eigenfunction expansion on an infinite arc in the upper or lower
complex half-plane, we get two distinct discrete expansions from the
isolated poles on the imaginary axis. A determination of the resulting
generalized functions shows that they are given by Wick rotated Landau
wavefunctions, including a Wick rotation of the background field. One
expansion corresponds to the Wick rotation $(x,E)\rightarrow (\i t,\i
E)$ and the other expansion to $(x,E)\rightarrow (-\i t,-\i E)$. This
shows that we can map one expansion to the other by a combined
time-reversal plus charge conjugation transformation
$\vec{C\,T}$. This suggests that the corresponding propagators relate
the propagation of charged particles and antiparticles in different
time directions respectively. We found an explicit expression for both
propagators in a background electric field alone, and determined
explicitly the appropriate domain for the 
expansion of noncommutative fields in these ``electric Landau
wavefunctions'' in terms of Gel'fand-Shilov spaces. 

However, a non-trivial result, which doesn't simply follow by Wick
rotation, is that stability of the theory requires the use of both
expansions simultaneously, i.e. we have to make the expansion in a
$\vec{C\,T}$-invariant way. This shows that in Minkowski spacetime we
effectively require twice as many degrees of freedom as compared to
the Euclidean case. This is most apparent in the matrix model
representation. While the Euclidean field theory is a one-matrix
model, the Lorentzian field theory is a two-matrix model.

This new matrix basis could now be used to implement the
renormalization programme for noncommutative field theory in Minkowski
spacetime. In the same way as the Landau basis was a crucial
ingredient in the proof of the renormalizability of some Euclidean
noncommutative field theories, the electric Landau basis could be used
in similar theories formulated in Minkowski spacetime. One could first
examine the Minkowskian version of original Grosse-Wulkenhaar
model~\cite{gw03}, which consists in adding an inverted harmonic
oscillator potential to the kinetic term of a real
$\phi^{\star4}$-theory, as given by the operator (\ref{D2Dt2}). The
corresponding propagator requires inversion of the analog of the
matrix appearing in (\ref{free2matrix}) at $\sigma=\frac12$, and can
be found using the techniques of~\cite{grvt05}. In this way our
formalism describes the appropriate analytic continuation of the
Grosse-Wulkenhaar models to Minkowski signature. Along these lines it
is interesting to explore the structure of the presumably inequivalent
quantization of the duality covariant field theory using the S-matrix
formalism in our two-matrix basis. It would also be
interesting to see if the exactly solvable self-dual matrix models of
Section~\ref{2matrixmodel} lead to any different nonperturbative
renormalizability properties compared to the Euclidean
case~\cite{lsz03}. All of these interesting renormalization issues are
left for future investigations.

We conclude by pointing out an interesting but somewhat unrelated
offspring of our analysis. A corollary of our work is a rigorous
mathematical proof of the electric-magnetic duality of the QED
effective action, which states that one can obtain the effective
action of charged particles in an electric background $E$ from that of
charged particles in a magnetic background $B$ by the substitution
$B\rightarrow\i E$~\cite{dh98}. The effective action is simply given
by
\begin{eqnarray}
 S_{\rm eff}&=&\i\log\Big(\,\int_\Phi \,\mathcal{D}\phi ~
\mathcal{D}\phi^*~\e^{\i S_0|_{\sigma=1}}\Big) \nonumber\\[4pt]
&=&-\i\log\,\det\left(\d^2+\mu^2\right) \nonumber \\[4pt]
&=&-\sum_{s=\pm}~\sum_{n=0}^\infty\,\i\log
\left(2\i s\, E\,(\mbox{$n+\frac12$})+\mu^2\right) \ ,
\end{eqnarray}
where we have omitted the infinite vacuum contribution in the second
line and used the generalized discrete spectrum in the third line. The
techniques developed in this paper may have further applications in
this context.

\subsection*{Acknowledgments}

We thank D.~Bahns, L.~Boulton, H.~Grosse, E.~Langmann, H.~Steinacker
and K.~Vogeler for helpful discussions. This work was supported in
part by the EU-RTN Network Grant MRTN-CT-2004-005104.

\setcounter{section}{0}

\appendix{\ \ \ Generalized eigenfunctions}\label{eigfunctions}

Given the relation with the inverted harmonic oscillator, it is
natural to introduce analogs of the standard ladder
operators. Defining
\begin{eqnarray}
 \vech{a}^\pm=\frac{1}{\sqrt{2\theta}}\,\big(\vech{p}\mp\vech{q}\big)
 \ , 
\label{ladderpm}\end{eqnarray}
one has the commutation relations
\begin{eqnarray}
 \big[\a^-\,,\,\a^+\big]=\i
\label{ladderalg}\end{eqnarray}
and the operator $\h$ can be represented as
\begin{eqnarray}
 \h&=&\mbox{$\frac{\theta}{2}$}\,\big(\a^+\,\a^-+\a^-\,\a^+\big) \ .
\end{eqnarray}
These operators are not ladder operators in the usual sense, since
they are not Hermitean conjugates of one another. Nevertheless, we can 
construct our basis distributions $\ket{f_n^\pm}$ and $\bra{f_n^\pm}$
in vacuum representations of the algebra (\ref{ladderalg}) defined
by applying these operators to states $\ket{0,\pm}$ and
$\bra{0,\pm}$, respectively, which are determined via the conditions
\begin{eqnarray}
 \a^{-s}\ket{0,s}&=&0 \ , \nonumber\\[4pt]
\bra{0,s}\a^{-s}&=&0 \ , \nonumber\\[4pt]
\bk{0,s}{0,-s}&=&1\label{klaus}
\end{eqnarray}
for $s=\pm$. The inner product in \eqref{klaus} follows again from the
fact that $\bra{f_n^\pm}=\bra{n}\v{\mp}^{-1}$ is only orthonormal to
$\ket{f_n^\mp}$. The generalized eigenstates
\begin{eqnarray}
 \ket{n,+}&:=&\frac{(-\i)^n}{\sqrt{n!}}\,\big(\a^+\big)^n\ket{0,+} \
 , \nonumber\\[4pt]
\ket{n,-}&:=&\frac{1}{\sqrt{n!}}\,\big(\a^-\big)^n\ket{0,-} \ ,
\nonumber\\[4pt]
\bra{n,+}&:=&\frac{\i^n}{\sqrt{n!}}\,\bra{0,+}\big(\a^+\big)^n \ ,
\nonumber\\[4pt]
\bra{n,-}&:=&\frac{1}{\sqrt{n!}}\,\bra{0,-}\big(\a^-\big)^n
\label{state4}\end{eqnarray}
have the desired properties
\begin{eqnarray}
 \h\ket{n,\pm}&=&\pm\i\theta\,\big(\mbox{$n+\frac12$}\big)\,\ket{n,\pm}
 \ , \nonumber \\[4pt]
\bra{n,\pm}\h&=&\mp\i\theta\,\big(\mbox{$n+\frac12$}\big)\,\bra{n,\pm}
\ , \nonumber \\[4pt]
\bk{n,\pm}{m,\mp}&=&\delta_{nm}\label{cond3} \ .
\end{eqnarray}
Consequently these states coincide (up to a phase factor) with the
distributions $\ket{f_n^\pm}=\v{\pm}\ket{n}$ and
$\bra{f_n^\pm}=\bra{n}\v{\mp}^{-1}$ constructed using the complex
scaling of Section~\ref{resexp1}.

The operators $\a^\pm$ together with the Weyl-Wigner correspondence 
now allow us to construct the generalized functions
$f_{n,m}^\pm(\vec{x})$ formally via 
\begin{eqnarray}
 f_{n,m}^\pm&=&\wi{\,\ket{n,\pm}\bra{m,\mp}\,}\=
\frac{(\mp\i)^n}{\sqrt{n!\,m!}}~\wi{\a^\pm}^{\star n}\star 
f_{0,0}^\pm\star\wi{\a^\mp}^{\star m} \ ,
\label{fnmladder}\end{eqnarray}
where $\wi{\a^s}^{\star n}$ denotes the $n$-fold star product
$\wi{\a^s}\star\cdots\star\wi{\a^s}$. With the notation $x^\pm=t\pm x$
and $\p{\pm}=\p{t}\pm\p{x}$, we find for an arbitrary function
$f(\vec{x})$ the star products
\begin{eqnarray}
 \wi{\a^\pm}(\vec{x})\star f(\vec{x})&=&\mbox{$\frac{\i}{2}$}\,\Big(
\mbox{$-\sqrt{\frac{\theta}{2}}~\p{\pm}\pm\i\sqrt{\frac{2}{\theta}}~
x^\mp$}\Big)f(\vec{x}) \ , \nonumber\\[4pt]
f(\vec{x})\star \wi{\a^\mp}(\vec{x})&=&\mbox{$\frac{\i}{2}$}\,
\Big(\mbox{$\sqrt{\frac{\theta}{2}}~\p{\mp}\mp\i\sqrt{\frac{2}{\theta}}~
x^\pm$}\Big)f(\vec{x}) \ .
\end{eqnarray}
This motivates the definition of new ``ladder operators'' on
$\vec{x}$-space given by
\begin{eqnarray}
 a_1^\pm\=\mbox{$\frac{\i}{2}\,\Big(-\sqrt{\frac{\theta}{2}}~
\p{\pm}\pm\i\sqrt{\frac{2}{\theta}}~x^\mp\Big)$} \qquad \mbox{and}
\qquad
a_2^\pm\=\mbox{$\frac{\i}{2}\,\Big(\sqrt{\frac{\theta}{2}}~
\p{\mp}\mp\i\sqrt{\frac{2}{\theta}}~x^\pm\Big)$} \ .
\label{a2}\end{eqnarray}
The new operators $a_i^\pm$, $i=1,2$ obey the nonvanishing commutation
relations $[a_i^-,a_j^+]=\i\delta_{ij}$, and with $\theta=2/E$ 
our basic differential operators can be expressed as
\begin{eqnarray}
 \d^2\=2E\,\big(a_1^+\,a_1^-+\mbox{$\frac\i2$}\big) \qquad \mbox{and}
\qquad \dt^2\=2E\,\big(a_2^+\,a_2^-+\mbox{$\frac\i2$}\big) \ .
\end{eqnarray}

The conditions \eqref{klaus} translated into this
language respectively give the differential equations
\begin{eqnarray}
 a_1^-f_{0,0}^+(\vec{x})\=a_2^-f_{0,0}^+(\vec{x})\=0 \qquad \mbox{and} \qquad 
a_1^+f_{0,0}^-(\vec{x})\=a_2^+f_{0,0}^-(\vec{x})\=0 \ ,
\label{DEf00}\end{eqnarray}
which can each be solved to give
\begin{eqnarray}
f_{0,0}^\pm(\vec{x})&=&\mbox{$\frac{\i E}\pi$}~\e^{\mp\i E\,(t^2-x^2)}
\end{eqnarray}
in the space $\scal'(\R^2)$, where the normalization constant has been
fixed by $\int_{\R^2}\, \dd\vec{x}~f_{0,0}^\pm(\vec{x})=1$.
\begin{lemma}
The generalized functions $f_{n,m}^{s,s'}=\wi{\,\ket{f_n^s}\bra{f_m^{-{s'}}}\,}$
vanish for distinct $s,s'=\pm$.
\label{moregenlemma}\end{lemma}
\begin{proof}
The analog of (\ref{DEf00}) for $f_{0,0}^{+,-}$ yields the two
differential equations 
\begin{eqnarray}
 \p{-}f_{0,0}^{+,-}(\vec{x})\=-\i E\,x^+\,f_{0,0}^{+,-}(\vec{x}) \qquad
 \mbox{and} \qquad
\p{-}f_{0,0}^{+,-}(\vec{x})\=+\i E\,x^+\,f_{0,0}^{+,-}(\vec{x}) \ ,
\end{eqnarray}
which together imply that $f_{0,0}^{+,-}=0$ by continuity. The same
argument leads to $f_{0,0}^{-,+}=0$. The result now follows from the
analog of (\ref{fnmladder}).
\end{proof}

We will now show that the explicit forms of the generalized
eigenfunctions $f_{m,n}^\pm(\vec{x})$ in Minkowski signature are simply
given by the Landau wavefunctions with Wick rotated parameters.
\begin{proposition}
The generalized eigenfunctions can be written as
\bea
f_{m,n}^\pm(z,\varphi)&=&\frac{|E|}{2\pi}\,\sqrt{\frac{n!}{m!}}~
(-1)^n~\e^{\mp\i E\,z^2/2}\,(\pm\i E)^{(m-n)/2}\,z^{m-n}~
\e^{\mp\,\varphi\,(m-n)}\,L_n^{m-n}\left(\pm\i
  E\,z^2\right)\nonumber\\ 
\label{f1}\\[4pt]
&=&\frac{|E|}{2\pi}\,\sqrt{\frac{m!}{n!}}~
(-1)^m~\e^{\mp\i E\,z^2/2}\,(\pm\i E)^{(n-m)/2}\,z^{n-m}~
\e^{\mp\,\varphi\,(m-n)}\,L_m^{n-m}\left(\pm\i E\,z^2\right) \ ,
\nonumber\\ 
\label{f2}\eea
where $z=\sqrt{t^2-x^2}$, $\varphi=\tanh^{-1}(x/t)$ and $L_n^k(y)$
are the associated Laguerre polynomials.
\label{fnmexplprop}\end{proposition}
\begin{proof}
We use the Wigner transformation formula (\ref{Groid}) and the
explicit form of the generalized eigenfunctions (\ref{fnpmfinal}) with the
electric field $E'=E/2=1/\theta$. Using the generating function for
the Hermite polynomials given by
\begin{eqnarray}
 \e^{\frac{-\xi+\xi\,
     q}{\mp\i\theta}}=\sum_{n=0}^\infty\,\frac{1}{n!}\,
\left(\frac{\xi}{\sqrt{\mp\i\theta}}\right)^n\,H_n
\big(q\big/\sqrt{\mp\i\theta}~\big) \ ,
\end{eqnarray}
we have
\begin{eqnarray}
 K^\pm(\xi,\eta;t,x)&:=&2\pi\,|\theta|\,\sum_{m,n=0}^\infty\,
\sqrt{\frac{2^{m+n}}{m!\,n!}}\,\left(\frac{\xi}{\sqrt{\mp\i\theta}}
\right)^m\,\left(\frac{\eta}{\sqrt{\mp\i\theta}}\right)^n\,
f_{m,n}^\pm(\vec{x}) \nonumber \\[4pt]
&=&\frac{1}{\sqrt{\mp\i\theta\,\pi}}\,\int_\R\, \dd k~
\exp\Big\{-\mbox{$\frac{1}{\mp\i\theta}$}\,\Big[\big(\xi^2-2\xi\,(t+k/2)
\big)-\big(\eta^2-2\eta\,(t-k/2)\big) \notag\\
&& \qquad \qquad \qquad \qquad \qquad
-\,\mbox{$\frac{1}{2}$}\,(t+k/2)^2-\mbox{$\frac{1}{2}$}\,(t-k/2)^2-
\i k\,x\Big]\Big\} \ . \label{peter}
\end{eqnarray}
Evaluating the formal Gaussian integral in \eqref{peter} gives finally the
generating function
\beq
K^\pm(\xi,\eta;t,x)=2\exp\left\{\frac{1}{\mp\i\theta}\,
\left[x^2-t^2+2\xi\,(t\mp x)+2\eta\,(t\pm x)-2\eta\,\xi\right]\right\}
\label{Kpmfinal}\eeq
in the space $\scal'(\R^4)$. The generalized functions
$f_{m,n}^\pm(\vec{x})$ can now be obtained by taking suitable
derivatives of (\ref{Kpmfinal}) with respect to the variables $\xi$
and $\eta$.

For $m\geq n$ one finds
\begin{eqnarray}
f_{m,n}^\pm(\vec{x})&=&\frac{1}{2\pi\,|\theta|}\,\frac1{\sqrt{m!\,n!}}\,
\left(\frac{\mp\i\theta}
{2}\right)^{(m+n)/2}\,\left.\frac{\partial^m}{\partial\xi^m}\,
\frac{\partial^n}{\partial\eta^n}K^\pm(\xi,\eta;t,x)\,\right|_{\xi=\eta=0}
\nonumber \\[4pt]
&=&\frac{\sqrt{m!\,n!}}{\pi\,|\theta|}~\e^{\frac{1}{\mp\i\theta}\,(x^2-t^2)}\,
\left(\frac{2}{\mp\i\theta}\right)^{(m-n)/2}\,(t\mp x)^{m-n} \nonumber
\\ &&\times\,\sum_{p=0}^n\,\left(\frac{2}{\mp\i\theta}\,\big(t^2-x^2
\big)\right)^{n-p}\,\frac{(-1)^p}{(m-p)!\,(n-p)!\,p!} \ .
\end{eqnarray}
This expression can be further simplified by introducing the rapidity
parameters $(z,\varphi)$ defined by
\begin{eqnarray}
t\=z\,\cosh\varphi \qquad \mbox{and} \qquad x\=z\,\sinh\varphi \ ,
\end{eqnarray}
such that $x^\pm=t\pm x=z~\e^{\pm\,\varphi}$. Introducing the summation
index $q=n-p$, inserting $E=2/\theta$ and using the definition of the
associated Laguerre functions
\begin{eqnarray}
 L_n^k(y)=\sum_{q=0}^n\,\frac{(n+k)!\,(-1)^q\,y^q}{(n-q)!\,(k+q)!\,q!}
 \ ,
\end{eqnarray}
we arrive at the explicit form for $m\geq n$ given by (\ref{f1}). The
calculation for $m<n$ is completely analogous and leads to
(\ref{f2}). However, using the identity~\cite[p.~321]{han75}
\begin{eqnarray}
(-1)^n\,r^{m-n}\,L_n^{m-n}\big(r^2\big)=(-1)^m\,r^{n-m}\,
L^{n-m}_m\big(r^2\big) \ ,
\end{eqnarray}
we see that both forms (\ref{f1}) and (\ref{f2}) are valid for
\emph{generic} $m,n\in\N_0$, and are thus equivalent.
\end{proof}

\appendix{\ \ \ Free two-point function}

In this appendix we will derive an explicit expression for the free
propagator (\ref{Cpmnm1},\ref{Cpmnm2}) at $\sigma=1$ in the spacetime coordinate
basis. We begin with a spectral expansion of the
propagator 
\begin{eqnarray}
 C_{\sigma=1}^\pm(\vec{x},\vec{x}'\,)&=&2\i\bra{\vec{x}}\big(\d^2+\mu^2
\big)^{-1}\ket{\vec{x}'\,}\=
\sum_{m,n=0}^{\infty}\,\frac{2\i f_{m,n}^\pm(\vec{x})\,f_{n,m}^\pm(
\vec{x}'\,)}{\pm\,\eps_m+\mu^2} \ ,
\end{eqnarray}
where we have used (\ref{Groid}). First we will evaluate the sum over
$n$. Substituting the expression \eqref{f1} for $f_{m,n}^\pm(\vec{x})$
and \eqref{f2} for $f_{n,m}^\pm(\vec{x}'\,)$ we get 
\begin{eqnarray}
 \sum_{n=0}^\infty\, f_{m,n}^\pm(\vec{x})\,f_{n,m}^\pm(\vec{x}'\,)
&=&\left(\frac{1}{\pi\,\theta}\right)^2~
\e^{\mp\i E\,(z^2+z'\,^2)/2}\,\frac{(\pm\i E\,z\,z'\,)^m}{m!}~
\e^{\mp\,(\varphi-\varphi'\,)\,m} \label{sumnfirst} \\
&&\times\,\sum_{n=0}^\infty \,n!\,\big(\pm\i E\,z\,z'~
\e^{\mp\,(\varphi-\varphi'\,)}\big)^{-n}\,L_n^{m-n}\big(\pm\i E\,z^2
\big)\,L_n^{m-n}\big(\pm\i E\,z'\,^2
\big) \ . \nonumber
\end{eqnarray}
Using the identity~\cite[eq.~(48.23.11)]{han75}
\begin{eqnarray}
 \sum_{n=0}^\infty\, n!\, c^n\,L_n^{m-n}(\xi)\,
L_n^{k-n}(\eta)=k!~\e^{c\,\xi\,\eta}\,(1-\eta\, c)^{m-k}\,c^m\,
L_k^{m-k}\big(\mbox{$\frac{(1-\xi\, c)\,(\eta\, c-1)}{c}$}\big)
\label{han1}\end{eqnarray}
for $k=m$, after a bit of algebra we find
\begin{eqnarray}
\sum_{n=0}^\infty\, f_{m,n}^\pm(\vec{x})\,
f_{n,m}^\pm(\vec{x}'\,)=\frac{1}{\pi^2\,\theta^2}~
\e^{\mp\i E\,(\vec{x}-\vec{x}'\,)^2/2}~
\e^{-\i z\,z'\sinh(\varphi'-\varphi)}\,
L_m\big(\pm\i E\,(\vec{x}-\vec{x}'\,)^2\big) \ ,
\end{eqnarray}
where the factor $m!\,c^m$ coming from \eqref{han1} cancels the same
factor appearing in the denominator of~(\ref{sumnfirst}). 

To obtain the full propagator we also have to carry out the summation
over the index $m$ to get 
\beq
C_{\sigma=1}^\pm(\vec{x},\vec{x}'\,)=\frac1{\pi^2\,\theta^2}~
{\e^{\mp\i E\,(\vec{x}-\vec{x}'\,)^2/2}~
\e^{-\i E\, z\,z'\sinh(\varphi'-\varphi)}}\,
\sum_{m=0}^\infty\,\frac{L_m\big(\pm\i E\,(\vec{x}-\vec{x}'\,)^2
\big)}{\pm\,E\,\big(\mbox{$m+\frac12\pm\frac{\i\mu^2}{2E}$}
\big)} \ .
\label{fullpropsumm}\eeq
Using the identity~\cite[eq.~(48.2.3)]{han75}
\begin{eqnarray}
 \sum_{m=0}^\infty\,\frac{1}{m+a}\,L_m^k(w)=
\frac{\Gamma(a)\,\Gamma(k)}{\Gamma(a-k)}~{}_1F_1(a;k+1;w)+
\Gamma(k)~{}_1F_1(a-k;1-k;w)
\label{han2}\end{eqnarray}
with $a=\frac12\pm\frac{\i\mu^2}{2E}$, $k=0$, $w=\pm\i
E\,(\vec{x}-\vec{x}'\,)^2$ and ${}_1F_1(a;b;w)$ a confluent
hypergeometric function, we finally obtain 
\begin{eqnarray}
C_{\sigma=1}^\pm(\vec{x},\vec{x}'\,)=
\pm\,\frac1{2\pi^2}\,{E~\e^{\mp\i E\,(\vec{x}-\vec{x}'\,)^2/2}~\e^{-\i E\,z\,z'
\sinh(\varphi'-\varphi)}}~{}_1F_1\big(
\mbox{$\frac12\pm\frac{\i \mu^2}{2E}$}\,;\,1\,;\,
\pm\i E\,(\vec{x}-\vec{x}'\,)^2\big)
\label{propspacetime}\end{eqnarray}
where we have again substituted $\theta=2/E$.
The domain of validity for the identities used above can be found
in~\cite{han75}, and applied in our instance as equalities in the
space $\scal'(\R^4)$. Note that the factor $\e^{-\i E\,
  z\,z'\sinh(\varphi'-\varphi)}$ breaks translation invariance, as
expected in an electric background.

\bibliographystyle{alpha}
\bibliography{UVIR_Mink6}

\end{document}